\documentclass[8pt,preprint,aas_macros]{aastex}

\usepackage{psfig}
\usepackage{rotating}

\newcommand{\xmm}{\textit{XMM-Newton}}
\newcommand{\rosat}{{\em ROSAT}}

\newcommand{\nh}{\mbox{$N_H$}}

\newcommand{\kteff}{\mbox{$kT_{\rm eff}$}}
\newcommand{\rinfty}{\mbox{$R_{\infty}$}}
\newcommand{\rns}{\mbox{$R_{\rm NS}$}}

\newcommand{\nuc}[2]{\mbox{$^{\rm #1}{{\rm #2}}$}}

\newcommand{\nhtt}{\mbox{$N_{H,22}$}}
\newcommand{\chisq}{\mbox{$\chi^2_\nu$}}
\newcommand{\Chisq}[3]{$\chi^2_\nu$/dof (prob.) = {#1}/{#2} (#3)}
\newcommand{\Msun}{\mbox{$M_\odot$}}
\newcommand{\Rsun}{\mbox{$R_\odot$}}
\newcommand{\Lsun}{\mbox{$L_\odot$}}

\newcommand{\xray}{\mbox{X-ray}}

\newcommand{\simlt}{\mathrel{\hbox{\rlap{\hbox{\lower4pt\hbox{$\sim$}}}\hbox{$<$}}}}
\newcommand{\simgt}{\mathrel{\hbox{\rlap{\hbox{\lower4pt\hbox{$\sim$}}}\hbox{$>$}}}}
\newcommand{\approxgt}{\mbox{$^{>}\hspace{-0.24cm}_{\sim}$}}
\newcommand{\approxlt}{\mbox{$^{<}\hspace{-0.24cm}_{\sim}$}}
\newcommand{\ee}[1]{\mbox{$10^{#1}$}}
\newcommand{\tee}[1]{\mbox{$\times 10^{#1}$}}
\newcommand{\ud}[2]{\mbox{$^{+ #1}_{- #2}$}}
\newcommand{\ppm}{\mbox{$\pm$}}

\newcommand{\unit}[1]{\mbox{$\rm\,#1$}}
\def\deg{\hbox{$^\circ$}}

\def\arcmin{\hbox{$^\prime$}}
\def\arcsec{\hbox{$^{\prime\prime}$}}
\def\sec{\mbox{$\,{\rm sec}$}}
\newcommand{\msun}{\mbox{$\,M_\odot$}}
\newcommand{\rsun}{\mbox{$\,R_\odot$}}
\newcommand{\lsun}{\mbox{$\,L_\odot$}}

\newcommand{\km}{\hbox{$\,{\rm km}$}}

\newcommand{\MeV}{\mbox{$\,{\rm MeV}$}}
\newcommand{\keV}{\mbox{$\,{\rm keV}$}}
\newcommand{\eV}{\mbox{$\,{\rm eV}$}}

\newcommand{\kpc}{\mbox{$\,{\rm kpc}$}}
\newcommand{\pc}{\mbox{$\,{\rm pc}$}}
\newcommand{\persec}{\mbox{$\,{\rm s^{-1}}$}}
\newcommand{\perksec}{\mbox{$\,{\rm ks^{-1}}$}}

\newcommand{\percmsq}{\mbox{$\,{\rm cm^{-2}}$}}
\newcommand{\percmcube}{\mbox{$\,{\rm cm^{-3}}$}}
\newcommand{\peryear}{\mbox{$\,{\rm yr^{-1}}$}}
\newcommand{\perval}[2]{{#1\mbox{$^{#2}$}}}
\newcommand{\cgsflux}{\mbox{$\,{\rm erg\,\percmsq\,\persec}$}}
\newcommand{\cgslum}{\mbox{$\,{\rm erg\,\persec}$}}
\newcommand{\cgsdensity}{\mbox{$\,{\rm g\percmcube}$}} 
\newcommand{\cgsaccel}{\mbox{$\,{\rm cm\,s^{-2}}$}} 

\def\OmCen{\mbox{$\omega$Cen}}
\def\sourcefour{XMMU~171433$-$292747}       %
\def\sourcefive{XMMU~171421$-$292917}       %
\def\sourceeight{XMMU~171413$-$293415}      %
\def\sourcenine{XMMU~171411$-$293159}       %
\def\TwoMassnine{2MASS~17141152$-$2931594}  %
\def\TwoMassfive{2MASS~17142095$-$2929163}  %

\begin{document}

\title{X-ray Spectral Identification of Three Candidate Quiescent
Low-Mass X-ray Binaries in the Globular Cluster NGC~6304}

\author{Sebastien Guillot\altaffilmark{1}, 
Robert E. Rutledge\altaffilmark{1}, 
Lars Bildsten\altaffilmark{2}, 
Edward F. Brown\altaffilmark{3}, \\
George G. Pavlov\altaffilmark{4}, 
and Vyacheslav E. Zavlin\altaffilmark{5}}
\altaffiltext{1}{Department of Physics, McGill University, 3600 rue
  University, Montreal, QC, H3A 2T8, Canada; guillots@physics.mcgill.ca, 
  rutledge@physics.mcgill.ca}
\altaffiltext{2}{Kavli Institute for Theoretical Physics and Department of 
  Physics, Kohn Hall, University of California, Santa Barbara, CA 93106; 
  bildsten@kitp.ucsb.edu}
\altaffiltext{3}{Department of Physics and Astronomy, National Superconducting 
Cyclotron Laboratory, and the Joint Institute for Nuclear Astrophysics, 
Michigan State University, 3250  Biomedical Physical Science Building, East 
Lansing, MI 48824-2320; ebrown@pa.msu.edu}
\altaffiltext{4}{The Pennsylvania State University, 525 Davey Lab, University 
Park, PA 16802; pavlov@astro.psu.edu}
\altaffiltext{5}{Space Science Laboratory, Universities Space Research 
Association, NASA MSFC VP62, Huntsville, AL 35805;
 vyacheslav.zavlin@nasa.msfc.gov}

\received{\today}
\slugcomment{Draft \today}

\begin{abstract}
We report the search for low-mass \xray\ binaries in quiescence
(qLMXBs) in the globular cluster (GC) NGC~6304 using \xmm\ observations.
We present the spectral analysis leading to the identification of
three candidate qLMXBs within the field of this GC,
each consistent with the \xray\ spectral properties of previously
identified qLMXBs in the field and in other GCs --
specifically, with a hydrogen atmosphere neutron star with radius
between 5--20\km.  One (\sourcefour, with $\rinfty
=11.6\ud{6.3}{4.6}$\,(D/5.97\kpc)\km\ and $\kteff=122\ud{31}{45} \eV$
is located within one core radius ($r_c$) of the centre of NGC~6304.
This candidate also presents a spectral power-law component
contributing 49 per cent of the 0.5--10\keV\ flux.  A second one
(\sourcenine\ with $\rinfty=10.7\ud{6.3}{3.1}$\,(D/5.97\,kpc)\,km and
$\kteff=115\ud{21}{16}$\,eV) is found well outside the optical core
(at $\sim32\,r_c$) but still within the tidal radius.  From spatial
coincidence, we identify a bright 2MASS infrared counterpart which, at
the distance of NGC~6304, seems to be a post-asymptotic giant branch
star.  The third qLMXB (\sourcefive\ with $\rinfty=23
\ud{69}{10}$\,(D/5.97\kpc)\km\ and $\kteff=70\ud{28}{20} \eV$) is a
low signal-to-noise candidate for which we also identify from spatial
coincidence a bright 2MASS infrared counterpart, with 99.916 per cent
confidence.  Three qLMXBs from this GC is marginally consistent with
that expected from the encounter rate of NGC~6304.  We also report a
low signal-to-noise source with an unusually hard photon index
($\alpha=-2.0\ud{1.2}{2.2}$).  Finally, we present an updated
catalogue of the \xray\ sources lying in the field of NGC~6304, and
compare this with the previous catalogue compiled from \rosat\
observations.
\end{abstract}

\keywords{Stars: neutron, \xray: binaries, Globular clusters:
individual: NGC~6304}

\maketitle


\section{Introduction}
Transiently accreting low-mass \xray\ binaries in quiescence (qLMXBs)
produce useful constraints on physical models of the interiors of
neutron stars.  They were first discovered during the post-outburst
periods of transient LMXBs Cen X-4 and Aql X-1, in which their faint
emission ($L_{\rm X}\sim10^{32}-10^{33}\cgslum$), interpreted as
thermal black-body emission, had emission areas a factor of 100
smaller than implied by the projected area of a 10\km\ neutron star
(NS).  Further observations of Cen X-4 and new observations of 4U
2129+47 \citep{garcia94} and 4U 1608$-$522 \citep{asai96b} confirmed
these observational properties of the class of qLMXBs.  At that time,
it was suggested that some low-level mass accretion on to the compact
object was required to power the observed luminosity, although what
set the low-mass accretion rate ($\dot{M}$) was unclear
\citep{jvp87,verbunt94}.

An alternate explanation to low-$\dot{M}$ accretion has become the
dominant explanation for the quiescent emission of qLMXBs \cite[BBR98
hereafter]{brown98}.  An up to date summary of the theoretical
developments of deep crustal heating and the related observations is
provided here.  In this different interpretation of the quiescent
thermal emission, the luminosity is provided not by accretion on to
the compact object, but by heat deposited in the neutron star crust by
pressure-sensitive nuclear reactions.  These reactions \citep[HZ08
hereafter]{haensel90,haensel03,gupta06,haensel08}, referred to as deep
crustal heating (DCH), occur as matter is piled at the top of the
neutron star surface, forcing the column below to greater density and
electron Fermi energies.  For example, beginning with pure
\nuc{56}{Fe} at the top of the crust, the increasing electron Fermi
energy induces electron capture at a density of $1.494
\tee{9}\cgsdensity$; because of the odd-even staggering of nuclear
masses, a second electron capture follows to produce \nuc{56}{Cr} and
releases 0.041\MeV\ per accreted nucleon (HZ08).  At a density
$1.114\tee{10}\cgsdensity$ the electron Fermi energy is again above
threshold for capture on to \nuc{56}{Cr}, and \nuc{56}{Ti} is produced
with an energy deposition of 0.036\MeV\ per accreted nucleon.  This
process continues, resulting in a series of electron captures, neutron
emission, and pycnonuclear reactions, through the depth of the crust,
until the matter reaches the density of undifferentiated equilibrium
nuclear matter, at which point it is no longer constituted of
differentiable nuclei.  In doing so, this reaction chain deposits
1.9\MeV\ per nucleon, distributed throughout the crust (HZ08).
Although the amount of heat deposited into the outer crust depends on
the composition of nuclei produced by the fusing of accreted H and He
\citep{gupta06}, the total amount of heat is dominated by neutron
emissions and pycnonuclear reactions in the inner crust, giving a
total heat deposition that is relatively insensitive to the nuclear
history of the accreted matter (HZ08).

In a steady-state, these reactions give rise to a time-average
luminosity, which is directly proportional to the time-average mass
accretion rate (BBR98):
\begin{equation}
  L = 9\tee{32}\,\frac{\langle \dot{M} \rangle}{10^{-11}\unit{\msun\,
    \peryear}}\,\frac{Q}{1.5\unit{MeV/amu}}\cgslum
  \label{eq:dch}
\end{equation}
\noindent where $Q$ is the average heat deposited in the NS crust per
accreted nucleon.  Of course, nuclear burning in the NS atmosphere
will produce not pure iron, but a wide range of nuclei
\citep{schatz99, schatz01}; this has been shown \citep{haensel03} to
possibly change $Q$ to range between 1.1-1.5\MeV\ per accreted
nucleon.  Moreover, the uncertain time-scale for the pycnonuclear
burning of \nuc{34}{Ne} \citep{yakovlev06}, which arises in the
Fe-chain, leaves uncertain whether that reaction is followed, or some
other reaction takes its place on a shorter time-scale, at slightly
greater density.  However, in the final analysis, the amount of heat
$Q$ deposited in the crust is roughly equal to the average mass defect
of nuclei at the top of the crust (near densities
$\approx\ee{9}\cgsdensity$) with undifferentiated equilibrium nuclear
matter at the bottom of the crust, near nuclear saturation density
$2.6\tee{14}\cgsdensity$ (HZ08).

Some qLMXBs exhibit significantly less photon luminosity than
predicted by Eq.~(\ref{eq:dch}). In several cases, such as
KS~1731$-$260 \citep{wijnands01b,rutledge02c,cackett06} and
1H~1905+000 \citep{jonker07b} the long-term mass-transfer rate is not
well constrained.  For the accreting millisecond pulsar SAX
J1808.4$-$3657, however, predictions of $\langle \dot{M}\rangle$ from
the integrated fluence agree with predictions based on the orbital
period and gravitational radiation losses \citep{bildsten01}. For this
system, the low quiescent luminosity is consistent with strong neutrino
emission from the core, such as from nucleon or hyperon direct Urca
processes \citep{heinke07}.

The energy source of the quiescent thermal luminosity aside, the
thermal spectrum can be explained (BBR98) as due to a realistic
neutron star hydrogen atmosphere model
\citep{zavlin96,rajagopal96,heinke06}, rather than a black-body
approximation imposed in previous works.  In the case of qLMXBs, the
atmosphere has been accreted from the low-mass post-main sequence
stellar companion (BBR98).  When accretion shuts off, metals will
settle gravitationally on a time-scale of $\sim$\,seconds
\citep{bildsten92}, leaving a pure H-atmosphere.  The emergent
photon spectra of H atmosphere NSs is physically different -- although
parametrically similar -- to that of blackbody thermal spectra
\citep{zavlin96}, which dramatically changes the derived emission area
radii from the $\approxlt 1\km$ derived from the black-body
assumption, to radii which are consistent with the entire area of the
neutron star \citep{rutledge99}.

The DCH luminosity/H atmosphere spectral interpretation has been applied
in the detection (and non-detections) of a large number of historically 
transient LMXBs, including: 
Cen X-4 \citep{campana00,rutledge01}; 
Aql X-1 \citep{rutledge01b} 
4U~1608-522 \citep{rutledge99};
4U~2129+47 \citep{rutledge00};
the transient in NGC~6440, SAX~J1748$-$2021 \citep{intzand01}; 
X1732-304 in Terzan 1 \citep{wijnands02}; 
XTE~J2123$-$058 \citep{tomsick04}; 
EXO~1747$-$214 \citep{tomsick05}; 
MXB~1659$-29$ \citep{cackett06}; 
1M~1716$-$315 \citep{jonker07a}; 
2S~1803$-$245 \citep{cornelisse07}; 
4U~1730$-$22 \citep{tomsick07}; 
1H~1905+000 \citep{jonker07b}.

The spectral analysis of qLMXBs with the H atmosphere interpretation
leads to the determination of \rinfty, the emission radius (also
called, the projected radius) of the neutron star.  It differs from
the physical radius by a factor $g_{r}$, the gravitational redshift:
\begin{equation}
  \rinfty = \frac{\rns}{g_{r}} = \rns  \left(1 - 
  \frac{2 G M_{\rm NS}}{\rns\ c^{2}} \right)^{-1/2}
  \label{eq:rinf}
\end{equation}
\noindent The effective temperature is also affected by the
gravitational redshift through the relation $T_{\rm
eff}^{\infty}=g_{r}T_{\rm eff}$.  A major uncertainty in accurate
measurements of \rinfty\ remains the distance $D$ to the system.
While high signal-toi-noise (S/N) \xray\ spectra are capable of
determining $\rinfty/D$ to statistical certainty, the 10--50 per cent
systematic uncertainties in the distances to field sources directly
contribute an additional 10--50 per cent uncertainty in \rinfty.  The
observational motivation for measuring \rinfty\ -- placing
observational constraints on the dense matter equation of state --
requires measurements of \rinfty\ to $\approx 5$ per cent accuracy, or
better \citep{lattimer04}.

The observational solution is to discover new qLMXBs, the distances to
which are known with greater certainty than the present group of field
sources.  Globular clusters (GCs) are the obvious place to search for
these, as they host an over-abundance of LMXBs and their distances are
precisely known or measurable.  Historically, it was suggested that
\xray\ binary systems in GCs are formed by capture from the remnants
of massive stars, i.e. neutron stars or black holes \citep{clark75}.
It was also noted that the ratio of \xray\ sources to the surrounding
stellar density is two orders of magnitude larger for GCs than for the
Milky Way.  For example, 12 out of $\sim$ 100 known bright \xray\
sources were lying within GCs while the population of GC
represents only $10^{-4}$ of the total mass of the Milky Way
\citep{hut92}.  The encounter rate is related to the physical
properties of GCs, so that the high stellar density in their core
creates a propitious environment for large encounter rates
\citep{verbunt02}.  Recently, a correlation was shown between the GC
encounter rate and the number of hosted \xray\ sources
\citep{pooley03} or between the GC encounter rate and the number of
hosted accreting NS binary systems \citep{heinke03b}.  In addition,
cluster metallicity can play a significant role in the number of LMXBs
in GCs \citep{kundu02,maccarone03,ivanova06,sivakoff07}.

To date, only a handful of qLMXBs in GCs capable of producing high S/N
\xray\ spectra are known: 
in Omega Cen \citep{rutledge02b,gendre03a};
X5 and X7 in 47~Tuc \citep{heinke03,heinke06}; 
in NGC~6397 \citep{grindlay01b}; 
in M13 \citep{gendre03b}; 
in M28 \citep{becker03}; 
A-1 in M30 \citep{lugger07}; 
in M80 \citep{heinke03c}; 
in M55 and NGC~3201 \citep{webb06}.
Others have been reported in the literature, but with lower S/N, or
high values of \nh, which makes them significantly less useful for the
primary motivation for searching for qLMXBs: providing measurements of
\rinfty.

In each case, the qLMXB was identified by its \xray\ spectrum alone,
consistent with a H-atmosphere neutron star with radius of $\approx
10\km$, without observation of an optical/IR counterpart, or evidence
of an historical \xray\ outburst.  In two cases -- X5 in 47 Tuc
\citep{edmonds02} and Omega Cen \citep{haggard04} -- the optical
counterpart of the binary was identified following the spectral
classification, lending support to the assertion that \xray\ spectral
classification can identify qLMXBs.

Some qLMXBs also exhibit a power-law (PL) spectral component which
dominates at photon energies above 2\keV.  The first detection of such
a component was made on the qLMXB Cen X-4 \citep{asai96b,asai98},
where the power-law contributes about 40 per cent of the 0.5--10\keV\
flux.  Observations of 47 Tuc (qLMXBs X5 and X7) also showed weak
power-law components (with photon index $\alpha=2.6{\rm -}3$)
\citep{grindlay01}.  However, analysis of more sensitive observations
concluded that PL components are not required \citep{heinke03}.  A
more recently identified candidate qLMXB in 47~Tuc, has shown a
power-law component that represents $65\ud{28}{22}$ per cent of the
0.5--10\keV\ unabsorbed flux \citep{heinke05}.  NGC~6440 have possible
qLMXBs with significant PL components, but the large absorption in the
direction of this GC prevents a justified classification of the \xray\
sources due to the low photon flux where the thermal spectrum is
dominant \citep{intzand01}.  In general, qLMXBs in GCs display no or
little power-law components while qLMXBs in the field more frequently
tend to have a significant power-law component.  A proposed
interpretation relates the power-law emission as the residual of a
recent accretion episode \citep{grindlay01}. An alternate explanation
evokes it to shock emissions via the emergence of a magnetic field
\citep{campana00b}.  However, neither of these two hypothesis are yet
supported by observational evidence.

A program of short-integration \xmm\ observations, using the pn
camera, have been undertaken to survey GCs for qLMXB candidates, to
increase the number of known qLMXBs in GCs.  This paper reports the
discovery of three candidate qLMXBs in NGC~6304 from spectral
classification, \sourcefour, \sourcenine\ and the low signal-to noise
source, \sourcefive.  These were found in short \xmm/pn exposures
targeted specifically for the discovery of examples of this class of
objects.  The organisation of this paper is as follows.  In
\S\ref{sec:obs} and \S\ref{sec:astrometry}, the data reduction, the
observations and the astrometric analysis are described.  In
\S\ref{sec:spectra}, the spectral analysis of the detected sources are
presented with detailed explanations for interesting sources,
including a hard source with the photon index $\alpha=
-2.0\ud{1.2}{2.2}$.  In \S\ref{sec:candidates}, the observational
properties of the three candidate qLMXBs are described.
\S\ref{sec:discuss} covers a comparison with a previous observation of
NGC~6304 with \rosat\ (\S\ref{sec:rosat}) and a comparison with the
expected number of qNS in GCs (\S\ref{sec:number}).  Finally,
\S\ref{sec:conclude} summarizes the results of this observation and
concludes.

\subsection{NGC~6304} 
\label{ngc6304}
The globular cluster NGC~6304 is located at the position
R.A.=17h14m32.1s and ${\rm Dec.}=-29\deg 27\arcmin 44.0\arcsec$,
(J2000) approximately 2.2\kpc\ from the galactic centre.  Its
integrated visual magnitude is $V=8.22$ and it has a color excess
$E\left(B-V\right)=0.53$.  NGC~6304 is moderately compact, with the
core radius $r_c=0.21\arcmin$, the half mass radius $r_{\rm HM}=
1.41\arcmin$ and the tidal radius $r_t=13.25\arcmin$.  It is a non
core-collapse GC with a concentration
$c=\log\left(r_{t}/ r_{c}\right)=1.80$ (assuming a King's profile) and
with a central luminosity density $\log\left(\rho_{0}\right)=
4.41\unit{\lsun\,pc^{-3}}$.  Finally, its metallicity $\left[{\rm
m/H}\right]$ is -0.56, implying Z = 0.00488, where $Z_{\odot} =
0.0177$ was used \citep{montalban04}.  All properties come from the
updated catalog of GCs \citep{harris96} or from a more
recent work \citep{valenti07}.  A more recent measurement of the
cluster distance modulus, $\left(m-M\right)_{0}= 13.88\pm0.03$,
provides an heliocentric distance of $5.97 \ppm{0.08} \kpc$
\citep{valenti05}.

\section{Data Reduction and Analysis}
\label{sec:red}
\subsection{Observation}
\label{sec:obs}

The target was observed on 4 Sept 2006 using \xmm.  The three \xray\
CCD cameras -- MOS1, MOS2, and pn -- were operating in imaging mode,
with exposure times of 12618\sec, 12626\sec\ and 11037\sec,
respectively.  All exposures were made with the medium filter.  The
present analysis is focussed on the pn camera \citep{xmmpn}, due to
its greater sensitivity in the low photon energy range (0.5--2.0\keV)
where qLMXBs are brightest.  However, a drawback of the pn instrument
is the multiple bad columns, which are accounted for during the
analysis.

\begin{figure}
  \centerline{~\psfig{file=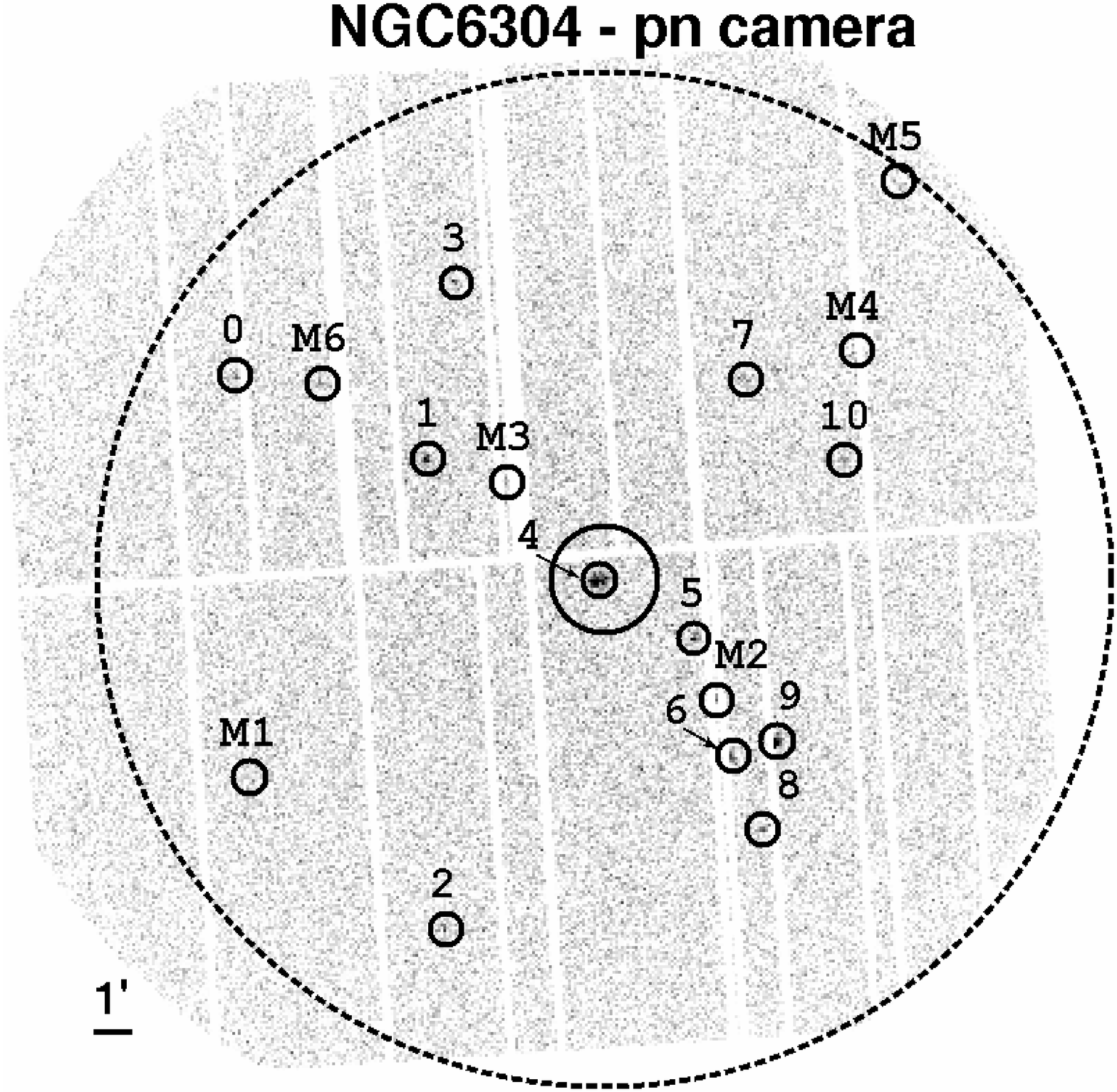,width=15cm,angle=0}~}
  \bigskip
  \caption[]{XMM/pn image of the globular cluster NGC~6304.  Each
    small circle shows the position of the 17 detected sources.  The
    small circles labelled M1 through M6 correspond to sources
    detected on the MOS cameras only.  The bigger solid circle and
    large dashed circle represent the half mass radius and the tidal
    radius (respectively) of NGC~6304.  The values $r_{\rm
    HM}=1.41\arcmin$ and $r_{\rm t} = 13.25\arcmin$ are from the
    Harris catalogue of globular clusters \citep{harris96}.  The
    source numbers correspond to those in Table~\ref{tab:all1} and
    Table~\ref{tab:allMOS}.
    \label{fig:image}}
\end{figure}

Data reduction was performed using \xmm\ Science Analysis System
v7.0.0, using standard analysis procedures.  The command {\tt epchain}
performs all required steps for preliminary data reduction.  The data
were checked for proton flares, but the standard analysis found none.
However, visual inspection of the exposure reveals a statistically
significant background variability during the observation.  Despite
this, the full observed time interval has been used for this analysis.
The spectral quality of the data is quantified using the task
{\tt epatplot}, as recommended before choosing the energy band
\footnote{From \xmm\ Science Operations Center
XMM-SOC-CAL-TN-0018,\cite{guainizzi08}}.  Over the whole field, {\tt
epatplot} shows a significant discrepancy below 0.45\keV\ between the
distribution of the various pattern events (single and double) as a
function of energy and the expected curves, therefore excluding the
use of this low energy band.  The 0.5--10\keV\ spectral energy range
was therefore chosen during the data reduction, the detection and the
subsequent steps of the analysis.  Only the event patterns recommended
by the \xmm\ Science team were used, that is, single and double
patterns for the pn camera, and single, double and quadruple patterns
for the MOS cameras.  An exposure map was then created for source
detection, which was performed with {\tt wavdetect} from CIAO v3.4\
\citep{ciao}.  Using the pn data, a total of 11 sources were detected
above a minimum {\tt wavdetect} relative exposure of 0.2 (see
Table~\ref{tab:all1} and Fig.~\ref{fig:image}).  Other {\tt wavdetect}
parameters used are the wavelet scales (1.0 2.0 4.0 8.0) and the
significance threshold ($3\tee{-6}$).  Using the same detection
parameters, 6 other sources are detected on a MOS1+MOS2 combined
image, most of them being coincident with bad columns on the pn
camera.  The source ID numbers in this paper (from 0 to 10) are given
by the default numbering of {\tt wavdetect}, attributing the ID
numbers by decreasing right ascension.  The additional MOS sources are
labelled M1 to M6.  The pn data was used to address the primary
purpose of this observation.  The source detected on the pn camera
were spectrally fitted in order to identify candidate qLMXBs.  For
completeness and as a secondary objective of creating a catalog of
\xray\ sources in the field of NGC~6304, the sources detected on the
MOS1+MOS2 image are reported in Table~\ref{tab:allMOS}.

\subsection{Astrometry}
\label{sec:astrometry}

The astrometric positions given in Table~\ref{tab:all1} were derived
in the following manner.  Astrometry was first obtained using {\tt
wavdetect} with the pn data, the MOS1 data, and the MOS2 data, all
separately.  Comparing the relative astrometry for sources detected by
each of these cameras, discrepancies are noticed in the relative
astrometry for some sources.  Sources \#5, \#6 and \#9 relative
distances from a reference source (see below) in the pn camera,
compared with that from the MOS cameras, were not consistent with the
statistical positional uncertainties.  The observed inconsistencies
seem to be due to bad columns on the pn camera, which are not
accounted for in the {\tt wavdetect} algorithm; thus {\tt wavdetect}
misestimates their positions in the direction opposite to the bad
column (which is the direction of the discrepancy in all three cases).

Therefore, combining astrometric data from the pn and MOS1 cameras
produced the \xray\ astrometric positions of the detected sources.
For most of the sources, the MOS1 data were used to deduce the
positions of the sources.  Some sources detected on pn, however, were
not detected on the MOS1 camera (\#2, \#3, and \#8).  The astrometric
positions of these sources are the pn position with a corrective
offset calculated from the difference in the relative astrometry of
the concerned source with source \#1 on the two cameras.  For example,
$RA^{\rm MOS1}_{3}=RA^{\rm pn}_{3} +\Delta RA^{\rm pn}_{\rm 3\,to\,1}$
is used to calculate the right ascension of source \#3 as it would be
on MOS1.  The reference \xray\ source is required to be detected on
both cameras and not adjacent to a bad column.  Only source \#1 and
\#4 fulfil these conditions in the pn camera data, but source \#4 was
discarded since it is possible that it corresponds to unresolved
multiple sources due to its presence in the core of NGC~6304.

Finally, an absolute astrometric correction -- into the frame of the
Two Micron All Sky Survey (2MASS) -- is applied through the
association of the \xray\ source \sourcenine\ with its counterpart
\TwoMassnine\ (99.988 per cent confidence, see
\S\ref{sec:sourcenine}), shifting the \xray\ positions by a change of
-1.66 arcsec in R.A. and -1.02 arcsec in Dec.  Previous to the frames
alignment, the separation between the pn and the 2MASS sources, 1.77
arcsec is consistent with the absolute astrometric uncertainty of
\xmm\ (2 arcsec within $1\sigma$\footnote{from \xmm\ Science
Operations Centre XMM-SOC-CAL-TN-0018, \cite{guainizzi08}}).  The
systematic uncertainty on the position, after the correction is
$1.2\unit{arcsec}\,\left( 1\sigma\right)$ \citep{jeffries06}.  Also,
the statistical uncertainty from the source detection (up to $\sim
1\unit{arcsec}$, depending on the source; see Table~\ref{tab:all1})
should be added in quadrature to the systematic error to obtain the
total uncertainty in position.

\subsection{Spectral Analysis}
\label{sec:spectra}

The \xray\ counts for spectra are extracted using a radius of 25
arcsec around each source, accounting for 81 per cent of the total
energy from the source at 1.5\keV.  The background counts are
extracted using an annulus centred at the position of the peak of the
exposure map (that is, the image centre), and with an average radius
equal to the distance between the source being analysed and the image
centre.  An annulus width of 40 arcsec is used for all sources, except
those sources close to the centre (sources \#4 and \#5) for which a
larger background area (100 arcsec circles) was extracted.  Around
each detected source, a circular area of 35 arcsec radius are excluded
from the background annulli.  A circular area of 35 arcsec excludes 87
per cent of the overlapping source counts (at 1.5\keV).  In all cases,
the sources contributes $\simlt 30$ per cent of the Poisson
uncertainty in the number of background counts -- to the background
regions.  This choice of background area addresses mirror vignetting,
which is the main cause of background variation across the field.  At
the source count rates encountered here, the effect of pile-up can be
neglected.  After the counts extractions, response matrices were
created using {\tt rmfgen} and {\tt arfgen}, rather than using the
predefined ones, in order to take into account the enclosed energies
and the extraction radii.

A threshold of 70 counts above the background was imposed before
considering the \xray\ source for a full spectral analysis.  Out of 11
detected sources, six satisfy this cut (ID \#1, 4, 5, 6, 9 and 10)
while the remaining five are too faint for a full spectral analysis
(ID \#0, 2, 3, 7 and 8).  The counts were binned into energy bins
between 0.5 and 10\keV\ with 25 counts per bins (or 20 counts per bins
for sources with less than 100 counts in excess of the background).

Using XSPEC v12.3 \citep{xspec}, the extracted spectra are fitted with a
tabulated neutron star hydrogen atmosphere model \citep{zavlin96}.
For this model, the radius is measured from the normalisation
parameter $\left( \rinfty / \left( d/10\kpc \right) \right)^{2} $,
assuming a distance $d=5.97\kpc$.  If this model provides a
statistically acceptable fit (Null hypothesis probability
$\approxgt10^{-2}$) and values of \rinfty\ and \kteff\ in the range of
previously observed qLMXBs (see Table~\ref{tab:NS}), $5-20\km$ and
$\sim 50-150\eV$ respectively, then, the source is classified as a
candidate qLMXB.  The spectral parameters are then confirmed with a
single-component model of neutron star hydrogen atmosphere included in
XSPEC (NSA model: {\tt nsa}, \cite{zavlin96}).  Alternatively, two
other NS atmosphere models are available in XSPEC: {\tt nsagrav}
\citep{zavlin96} and {\tt nsatmos} \citep{mcclintock04,heinke06}.
When using any of those models, the mass is kept constant at 1.4\msun\
and the model normalisation parameter ($1/d^2$) is set assuming
$d=5.97\kpc$.  The {\tt nsa} model contains a magnetic field parameter
that is set to zero here while {\tt nsagrav} and {\tt nsatmos} assume
non-magnetic atmospheres.  In all cases, the models {\tt nsa}, {\tt
nsagrav}, and {\tt nsatmos} give consistent parameters.  Regarding the
surface gravity $g$, {\tt nsa} has been implemented with a fixed value
$g=2.43\tee{14}\cgsaccel$, {\tt nsagrav} has been computed for a range
$g=(0.1-10)\tee{14} \cgsaccel$ and {\tt nsatmos} in the range
$g=(0.63-6.3)\tee{14} \cgsaccel$

In the case where the NSA tabulated model provides a non-acceptable
fit, a visual inspection and a F-test indicate whether a photon
power-law should be added to the spectral fit.  If a significant
portion of the high energy tail is in excess of the NS atmosphere fit,
a simple power-law is added.  This is justified by the fact that
qLMXBs sometimes display a hard power-law component which dominates
the spectrum above 2\keV.  If the fit is still not acceptable, the
spectrum is fitted with a power-law alone and the flux is obtained.
For low signal-to-noise ratio (S/N) source, the slope of the power-law
is fixed at $ \alpha = 1$ to estimate the flux.  For all model fits,
the photo-electric galactic absorption (the multiplicative model {\tt
wabs} in XSPEC) with a fixed value of $\nh=0.266\tee{22}
\unit{atoms\percmsq}$ (written $\nhtt=0.266$ afterwards) at the
position of the NGC~6304\footnote {From
http://cxc.harvard.edu/toolkit/colden.jsp using the NRAO data
\cite{dickey90}} is included in the spectral fit.

A major technical problem arises during the spectral fitting with the
{\tt nsa} model, as well as with the {\tt nsagrav} and {\tt nsatmos}
models.  All three of them were implemented in {\tt XSPEC} with
restrictive parameter spaces.  In particular, the radius parameter
space is defined only between 5\km\ and 20\km\ for {\tt nsa}, between
6\km\ and 45\km\ for {\tt nsatmos} and between 6\km\ and 30\km\ for
{\tt nsatmos}.  In this case, while the best-fit may be statistically
acceptable, and the parameter values obtained are consistent with
those expected for a qLMXB, the parameter uncertainty range is greater
than that tabulated in the {\tt XSPEC models} and the uncertainty
calculation fails due to zero diagonal components in the parameters
matrix.  The consequence of the imposed cut off in the parameter space
would be a misrepresentation of the parameter uncertainties,
preventing any conclusions to be drawn from the spectral fit.  In
those cases, the tabulated model \citep{zavlin96} and the parameters
resulting from the fit are used, since they do not suffer from this
constraint.  In all cases of identified qLMXBs, the MOS1 and MOS2
spectra are extracted in a similar manner as described above, and used
those to perform a simultaneous fit.  This procedure reduced the
uncertainties on the temperature and radii of the candidate qLMXBs, by
up to fifty per cent.

The results of this spectral analysis are given in Table
\ref{tab:all2}.  The following subsections present two sources that
are not candidates qLMXBs but which were above the signal-to-noise
cut-off for spectral analysis and the results of the spectral fitting
of source \#8, an \xray\ source with an extremely hard photon
power-law.  Finally, the spectral results for the three qLMXB
candidates are described in \S\ref{sec:candidates}.

\subsubsection{XMMU~171516-292224 - Source \#0}
This source, located far from the optical centre of the GC ($\sim 52
r_{c}$), is not spectrally consistent with a qLMXB at the distance of
NGC~6304.  Using the tabulated H-atmosphere model with a fixed \nh
($\nhtt = 0.266$), gives a non acceptable fit
(\Chisq{3.0}{12}{3\tee{-4}}) to the spectrum of the source.  Allowing
\nh\ to vary provides an acceptable fit \Chisq{1.2}{11}{0.28} with
$\nhtt = 18.2\ud{11.7}{9.4}$ but the resulting projected radius
$\rinfty = 1.3\ud{0.25}{0.22}\km$ excludes a possible quiescent
neutron star interpretation since the value is significantly smaller
than previously measured for such objects.

A power law spectral model is therefore chosen to characterize the
\xray\ emission and it is found that with $\nhtt = 0.266$, the photon
index is $\alpha = -0.24\ud{0.62}{0.82}$ for which the statistic is
\Chisq{1.0}{12}{0.45}.  If the absorption is chosen to be a free
parameter, the fit is still statistically acceptable
(\Chisq{0.84}{11}{0.60}) with the parameters $\nhtt =
1.6\ud{1.6}{6.2}$ and $\alpha = 0.3\ud{2.4}{1.0}$.  Finally, a search
for a possible 2MASS counterpart found no source within 3 sigma of the
position of source \#0.

\subsubsection{XMMU~171453-292436 - Source \#1}
Using the H-atmosphere tabulated model, a non acceptable fit with
\Chisq{2.47}{15}{0.001} is obtained when \nh\ is fixed at $\nhtt =
0.266$.  On the other hand, as \nh\ is allowed to vary, the best fit
(\Chisq{0.98}{14}{0.47}) projected radius $\rinfty =
0.28\ud{0.11}{0.08}\km$ excludes the qLMXB classification for this
source.  In addition, the best-fit (with fixed \nh) exhibits an excess
high energy tail, suggesting an additional power-law component at high
energies.  Adding a power-law component to the assumed spectrum
results in a dominating power-law ($\sim 100$ per cent of the source
0.5--10\keV\ flux) of slope $\alpha \sim 0.5$.  The NS atmosphere
component is therefore abandoned and the spectral fit is restricted to
a single power-law for which the best-fit photon index is
$\alpha=0.52\ud{0.34}{0.35}$ with a value \Chisq{0.96}{15}{0.5} in the
case where $\nhtt=0.266$.  However, as $\nhtt=2.3\ud{1.9}{1.6}$, the
photon index is $\alpha=1.7\ud{1.4}{0.9}$ for which
\Chisq{0.65}{14}{0.82}.

\subsubsection{XMMU~171417-293222 - Source \#6}
The NSA tabulated model produces an acceptable fit
(\Chisq{1.3}{12}{0.21} for $\nhtt = 0.266$ and \Chisq{1.3}{11}{0.21}
for $\nhtt = 20.8\ud{108}{20.8}$) but in both cases the obtained
parameters ($\rinfty<1\km$ and $\kteff \sim 0.5\keV$) are not in
accordance with expected values for NSs in quiescence.  Also, visual
inspection of the fitted spectrum indicates a significant high energy
tail.  The fit of the model (tabulated model{\tt+PL}) is also
statistically acceptable (\Chisq{0.96}{10}{0.48}) but, similarly to
source \#1, $\sim 100$ per cent of the unabsorbed flux comes from the
power-law component.  The single power-law model ($\alpha
=0.64\ud{0.87}{0.90}$) provide a statistically acceptable fit
(\Chisq{0.75}{12}{0.70}).  The spectrum fit with a fixed slope
($\alpha=1$) results in an almost unchanged \chisq\ statistic
(\Chisq{0.75}{13}{0.71}).  Finally, as both the absorption and the
photon index are allowed to vary, the best fit parameters are $\alpha
=0.45\ud{1.0}{0.8}$ and $\nhtt\leq0.9$ for \Chisq{0.78}{11}{0.66}.

\subsubsection{\sourceeight\ - Source \#8}
\label{sec:sourceeight}
Even though it is one of the faint sources, \sourceeight\ is
interesting since a single power-law spectral model suggests a
spectrally hard source.  As \nh\ is kept fixed at $\nhtt = 0.266$, the
best fit photon index $\alpha = -2.0\ud{1.2}{2.2}$ with
\Chisq{1.726}{7}{0.10} implying an unabsorbed flux $F_{X}=2.8\tee{-13}
\cgsflux$.  When the absorption is a free parameter, the photon index,
$\alpha=-2.1$, is similar to the case where \nh\ is fixed , but the
large \chisq\ value, \Chisq{2.02}{6}{0.06}, greater than 2 does not
allow for error estimation.

\begin{figure}
  \centerline{~\psfig{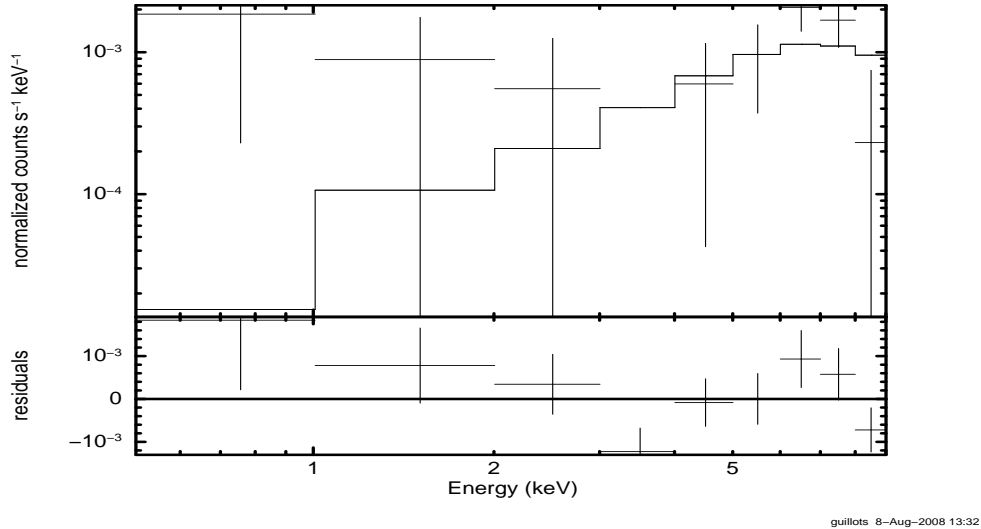}~}
  \bigskip
  \caption[]{Folded spectrum of source \#8 (\sourceeight) fitted with an absorbed
    power law model ($\alpha=-2.0\ud{1.2}{2.2}$) with the H-absorption
    fixed $\nhtt=0.266$.  The binning of the counts is different from
    the standard binning (minimum of 20 or 25 counts per bin).  Due to
    the low signal-to-noise of this source, bins of 1\keV\ width
    (except for the 0.5 to 1\keV\ bin) are chosen.
    \label{fig:Obj8Pow}}
\end{figure}

\begin{figure}
  \centerline{~\psfig{file=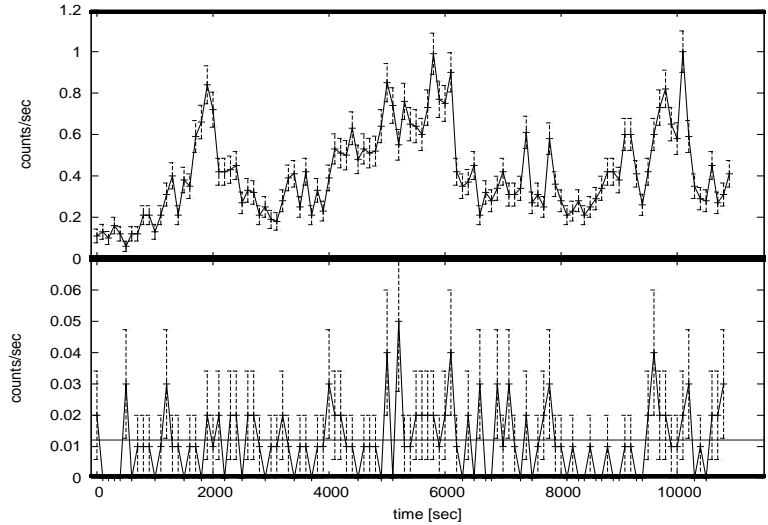,width=11.0cm,height=7cm,angle=0}~}
  \bigskip
  \caption[]{Lightcurve between 5 and 10\keV, with 100\sec\ bins.
    Bottom: lightcurve of extraction circular region of source \#8 (25
    arcsec).  The horizontal line at 0.0121 counts/sec (0.0078
    background counts/sec expected) is the mean count rate.  Top:
    lightcurve of the background area chosen for source \#8.
    \label{fig:Obj8LC}}
\end{figure}

Most of the counts are in the range 5--10\keV\
(Fig.~\ref{fig:Obj8Pow}).  The 5--10\keV\ light curve (100\sec\ bins)
exhibits no variability (Fig.~\ref{fig:Obj8LC}).  The average number
of counts per bin is 1.12, and 0.78 counts are expected on average due
to the background.  So, the average number of source counts is 0.43.
A bin with 5 counts is found and the Poisson probability of finding a
bin with such a peak in the lightcurve with a mean of 1.21 is 2.2 per
cent.  Also, the peak does not represent a variation by more that a
factor of 10 of the mean number of source counts.  The spectrally hard
source is consistent with constant emission throughout the duration of
the observation, to within a factor of $<10$ in any 100 second bin.
This excludes a source type such as an X-ray flash or soft gamma-ray
repeater (SGR), which tend to be spectrally hard in the 1--10 keV
energy band as exhibited by this object, but exhibit transient
emission of duration $\sim$few msec, to $\sim$100 sec.  No detailed
variability analysis can be performed due to the highly variable
background.  In conclusion, no variability is detected on a $\approxlt
100\sec$ time-scale.

To confirm the existence this source, the MOS2 camera data were used
(the CCD 6 on MOS1 containing source \#8 was inoperative during the
observation).  The same data reduction as described for the pn camera
and the same detection procedure is performed.  As explained before,
the data quality of the MOS camera is much better than the pn but its
sensitivity is not as good as the pn camera.  On MOS2, a total of 73
counts are detected within 25 arcsec of the expected position of
source \#8, with $2254\pm47.5$ counts in the background annulus.  From
the scaling factor between the source and the background areas,
$50.3\pm1.1$ background counts are expected within the source
extraction region.  Therefore, there is an excess of $22.7\pm8.6$
counts, corresponding to a low-significance detection ($2.6\sigma$); the
corresponding flux, assuming a power-law of slope $\alpha=-2.0$, is
$2.6\tee{-13}\cgsflux$ using the tool {\it webPIMMS}
\footnote{available at http://heasarc.nasa.gov/Tools/w3pimms.html},
which is consistent with the 2.8\tee{-13}\cgsflux\ value measured from
the pn spectrum.  Cross identification with the 2MASS and Digital Sky
Survey (DSS) catalogues did not exhibit any counterparts in the J and
B bands within 3$ \sigma $ of the X-ray position, respectively.

While the low signal-to-noise ratio of this detection (S/N = 4.0) does
not permit any detailed analysis of this source, it is unusually
spectrally hard for an object detected in the 0.5--10\keV\ band.
Examination of the unfiltered event list (i.e. including all event
patterns) finds no unusual number of rejected events, which does not
support an interpretation of the source detection as due to a particle
impact on the detector or an instrumental effect such as a transient
thermal excess on the detector. Higher sensitivity observations are
required to better characterize this source.

\section{Candidate qLMXBs}
\label{sec:candidates}

This section presents the spectral analysis of sources which, on the
basis of this spectral analysis, are candidate qLMXBs.

\subsection{\sourcefour\ - Source \#4}
\label{sec:sourcefour}

This \xray\ source, located near the centre of the cluster, has the
highest count rate among the sources in the field
($43.8\pm2.8\unit{cts\perksec}$).

\begin{figure}
  \centerline{~\psfig{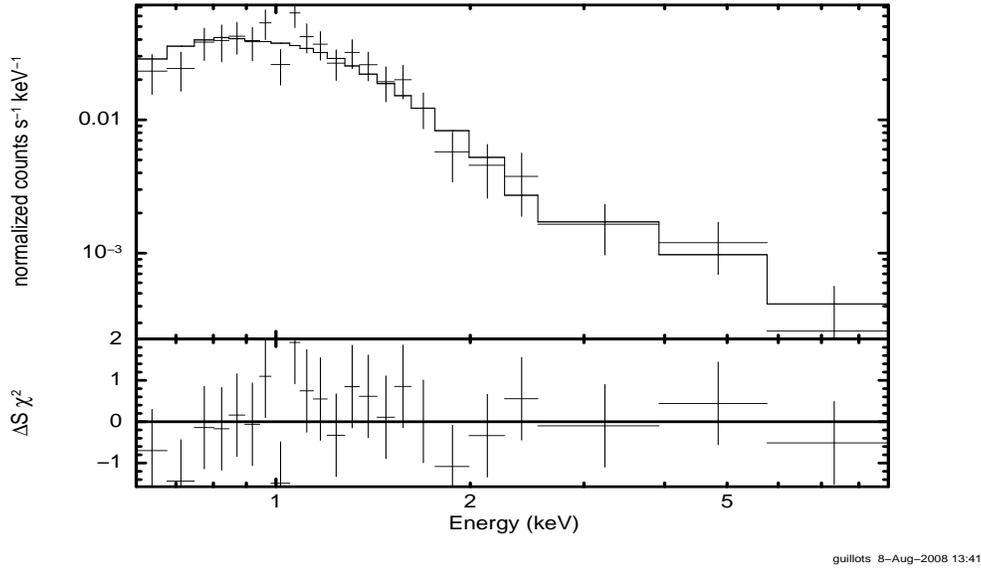}~}
  \bigskip
  \caption[]{Folded model to the pn camera spectrum of source \#4
    (\sourcefour).  The solid line is the best fit model of a neutron
    star atmosphere {\tt nsa} ($\nhtt=0.266$ held fixed) together with
    a power-law component of slope $\alpha=1.2\ud{0.7}{0.8}$.  Due to
    the low signal-to-noise at large energies (above 2.5\keV), the
    last 9 bins are grouped into three separate bins.
    \label{fig:Obj4NsaPow}}
\end{figure}

The fit with the tabulated NS atmosphere model is statistically
acceptable (\Chisq{1.24}{21}{0.20}) and is producing values for
\rinfty\ and \kteff\ in the range expected for a NS in quiescence.
However, there exists a systematic excess of counts above the best-fit
model at photon energies larger than 3\keV\, suggesting an additional
high-energy spectral component, which are accounted for by the
addition of a power-law.  The low probability value of the F-test
(prob=0.014) suggests that adding a power-law component better
describes the data than the absorbed H-atmosphere alone.  The best-fit
(tabulated) NSA model with the power-law component has a fit statistic
\Chisq{0.69}{19}{0.83} (see Fig.~\ref{fig:Obj4NsaPow}).  Also,
$\rinfty=8.1\ud{8.3}{2.5}\km$ and $\kteff = 127\ud{31}{29}\eV$ with
$\nhtt = 0.266$.  These values are in agreement with other known
qLMXBs (Table~\ref{tab:NS}).  However, when leaving the absorption
free $\nhtt=0.51\ud{0.18}{0.30}$ for this model, the fit is acceptable
(\Chisq{0.63}{18}{0.88}) but the radius is larger than expected for a
neutron star, $\rinfty=47\ud{190}{21}\km$.

The fit with fixed \nh\ is confirmed with a {\tt nsa+powerlaw} model,
for which the \chisq\ statistic is \Chisq{0.83}{19}{0.67}.  The radius
($\rns\ = 8.3\ud{9.2}{3.0}\km$) and temperature ($\kteff =
117\ud{59}{44}\eV$) suggests a candidate qLMXB since the \xray\
spectrum is consistent with the thermal spectrum of observed qLMXBs at
the distance of NGC~6304.  The slope of the power-law is $\alpha =
1.5\ud{0.9}{0.8}$ and this component of the model contributes to 49
per cent of the total unabsorbed 0.5--10\keV\ flux of the source.  A
simultaneous fit using the MOS1, MOS2 and pn spectra improved the
statistics (\Chisq{0.85}{42}{0.75}) and the uncertainties on the
parameters.  The radius and temperature of this candidate qLMXB are
now $\rns\ = 8.1\ud{4.2}{2.4}\km$ and $\kteff = 122\ud{31}{45}\eV$,
and the photon index is $\alpha = 1.2\ud{0.7}{0.8}$.  The value \rns\
is converted to \rinfty\ assuming a 1.4\msun\ neutron star using
Eq.~\ref{eq:rinf}: $\rinfty = 11.6\ud{6.3}{4.6}\km$.

The absence of significant intensity variability over the time scale
of the observation further supports the classification of this source
as a qLMXB.  Moreover, the luminosity is consistent with being the
same as that observed during the \rosat\ observation 14 yr earlier
(see Table \ref{tab:CompFluxB} and section \ref{sec:rosat}).  This is
consistent with the expected stable thermal luminosity from a qLMXB on
this time-scale \cite[BBR98,][]{ushomirsky01}.

The possibility of two unresolved sources -- one spectrally hard and
one spectrally soft -- is investigated by separating the data into
soft ($<1.5\keV$) and hard ($>1.5\keV$) images and performing a source
detection as described in Sec.~\ref{sec:obs}.  Source \#4 is detected
in both bands, $24.6\sigma$ and $4.5\sigma$ significances in the soft
and hard bands respectively (the 1.5\keV\ division was selected as the
lowest possible photon energy which provides a significant source
detection -- a cutoff of 2.0\keV\ finds no source with $>3\sigma$
significance above 2.0\keV).  A positional offset of $3\pm0.9$ arcsec
is measured between the positions of the source in the two images,
which is marginally consistent with a single source.  It is possible
that an unresolved hard \xray\ source lie close to the candidate qLMXB
and was interpreted as a high energy tail in the spectrum of source
\#4.  In other words, it is not possible to exclude that higher
spatial resolution observations may resolve \sourcefour\ into multiple
\xray\ sources.

\subsection{\sourcenine\ -  Source \#9}
\label{sec:sourcenine}

\subsubsection{Spectral Characterisation of the \xray\ source}
\begin{figure}
  \centerline{~\psfig{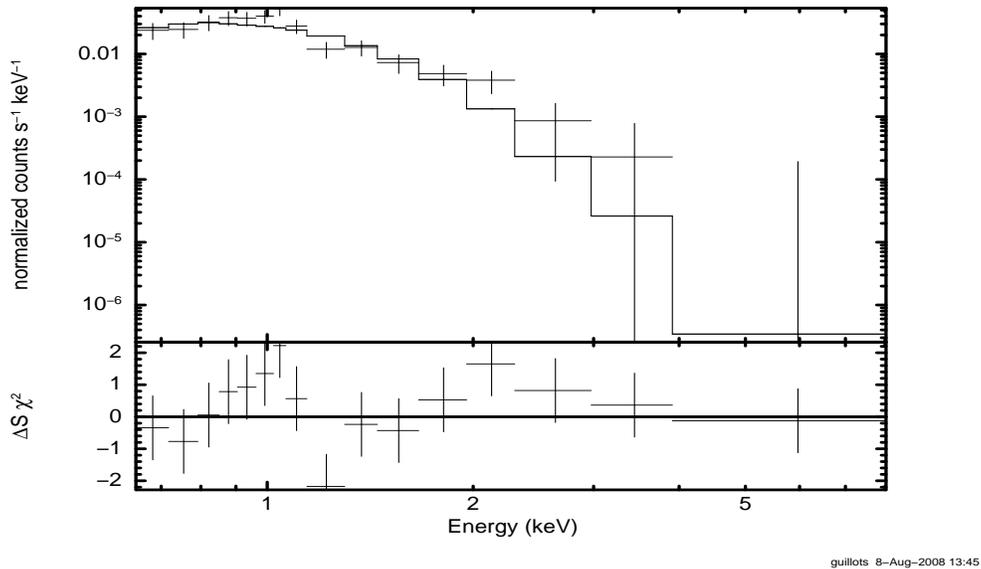}~}
  \bigskip
  \caption[]{Folded model to the pn camera spectrum of source \#9
    (\sourcenine).  The solid line is the best fit (tabulated) model
    of a NS H-atmosphere ($\nhtt=0.266$ held fixed).  Due to the low
    signal-to-noise at large energies, the last three bins are grouped
    into a single bin above 4\keV.
    \label{fig:Obj9folded}}
\end{figure}

The fit with the tabulated NS atmosphere model with fixed \nh is
statistically acceptable and does not require an additional power-law
(Fig.~\ref{fig:Obj9folded}).  The \chisq\ statistic is
\Chisq{1.29}{16}{0.20}, the projected radius is $\rinfty =
15.3\ud{15.5}{5.2}\km$ and the effective temperature is $\kteff =
100\ud{24}{19}\eV$.  Letting the absorption vary also gives a
statistically acceptable fit (\Chisq{1.38}{13}{0.16}) with the
following parameters: $\nhtt = 0.20\ud{0.37}{0.20}$, $\rinfty
=9.8\ud{31}{0.9}\km$ and $\kteff =115\ud{91}{61}\eV$, all
statistically consistent with the fit with a fixed absorption.

The confirmation with any of the models {\tt nsa, nsagrav} and {\tt
nsatmos} fails because the fit, while statistically acceptable,
produces error regions larger than the allowed parameter space in {\tt
XSPEC}.  It is therefore not possible to derive the uncertainty region
for the spectrum of \sourcenine\ using these models.  In consequence,
the results of the tabulated model show that the model fit is
consistent with the thermal spectrum of a qLMXB at the distance of the
NGC~6304. Moreover, no significant variability over the time-scale of
the observation is observed, with weak constraints.  To better improve
the statistics and uncertainties, the MOS1, MOS2 and pn spectra are
fitted simultaneously.  The resulting fit is statistically acceptable
(\Chisq{1.20}{31}{0.21}) and the obtained parameters are: $\kteff =
115\ud{21}{16}\eV$ and $\rinfty = 10.7\ud{6.3}{3.1}\km$.  Those values
for \sourcenine\ are reported as a candidate qLMXB in
Tables~\ref{tab:all2} and \ref{tab:NS}.

For completeness, additional spectral fits are provided.  First, A
single power law with fixed absorption (\nhtt = 0.266) gives an
acceptable fit (\Chisq{1.37}{14}{0.16}) with a soft photon index,
$\alpha = 3.51\ud{0.40}{0.37}$.  Leaving the galactic absorption \nh\
free $\nhtt = 0.57 \ud{0.44}{0.17}$, a steeper power-law is found
$\alpha = 5.0\ud{2.4}{1.5}$ with a \chisq-statistic
\Chisq{1.28}{13}{0.21}.  A single temperature Raymond-smith plasma fit
with variable absorption and solar metallicity also produces an
acceptable fit (\Chisq{1.03}{13}{0.42}) with the following best-fit
parameters: $\nhtt = 0.19\ud{0.14}{0.15}$ and
$\kteff=0.76\ud{0.26}{0.17}\keV$.  Thus, the spectrum of this (highest
S/N) X-ray source is consistent with other spectral interpretations;
however, such a steep power-law (either $\alpha=3.5$, or $\alpha=5.0$)
is usually interpreted as indicating a thermal spectrum.

As mentioned above, the pn camera suffers bad pixel quality.  Two of
the detected sources (\#6 and \#9) are overlapping with bad columns.
So for \sourcenine\ , an additional analysis using the MOS 1 and 2
cameras is performed.  Source \#9 is detected with a signal-to-noise
ratio of 17.7 and 17.6 for MOS1 and MOS2 respectively.  Using the
tabulated model, the spectral fits of the MOS1 and MOS2 data and the
simultaneous spectral fit of the MOS1, MOS2 and pn data are all
statistically acceptable and give results similar to the ones obtained
from the pn camera alone.

\subsubsection{Search for an IR counterpart}
\label{sec:searchIRC}
Using the online 2MASS Point Source Catalogue (2MASS--PSC), a probable
IR counterpart (\TwoMassnine ; $m_{J}=8.796(22)$, $m_{H}=8.361(18)$,
$m_{K}=8.213(20)$, where the numbers in parenthesis are the 1$\sigma$
uncertainties in the preceding digits) is found located at a distance
of 1.77 arcsec (before astrometric correction to the 2MASS frame).
For comparison, there are 6 stars of this magnitude or brighter in an
annulus ($r_{\rm in}=4.5\arcmin$ and $r_{\rm out}=8\arcmin$) about the
centre of the GC.  The probability that another source as bright or
brighter lies as close or closer to our \xray\ source is 0.012 per
cent.  Therefore, the association between \sourcenine\ and
\TwoMassnine\ is identified on the basis of spatial proximity, with
99.988 per cent confidence. Also, the USNO-B1.0 catalogue
\citep{usnob1} lists an object at this location (separated by $<$0.5
arcsec) with $B=11.66$ and $B=11.28$. The two different magnitudes may
indicate time variability of the counterpart.

A photometric study of NGC~6304 finds an observed $V=10.65$ for
\TwoMassnine\ (S. Ortolani, private communication, 2008); this is
about 3 magnitudes above the observed tip of the red giant branch
\citep{ortolani00} (see Fig~\ref{fig:VBV}), however, due to saturation
of the CCD, the photometry of very bright stars, like this
counterpart, was not performed -- thus, objects as bright as this were
not included in the published data\footnote{The photometry for
\TwoMassnine\ was performed by S. Ortolani from a separate shorter
(10\sec) V-band exposure}.  Inspecting by eye a $V$-band exposure of
NGC~6304 (S. Ortolani, private communication, 2008) shows that
$\sim$15--20 such saturated stars were not included in the $BV$
photometric study of this GC.

\begin{figure}
  \centerline{~\psfig{file=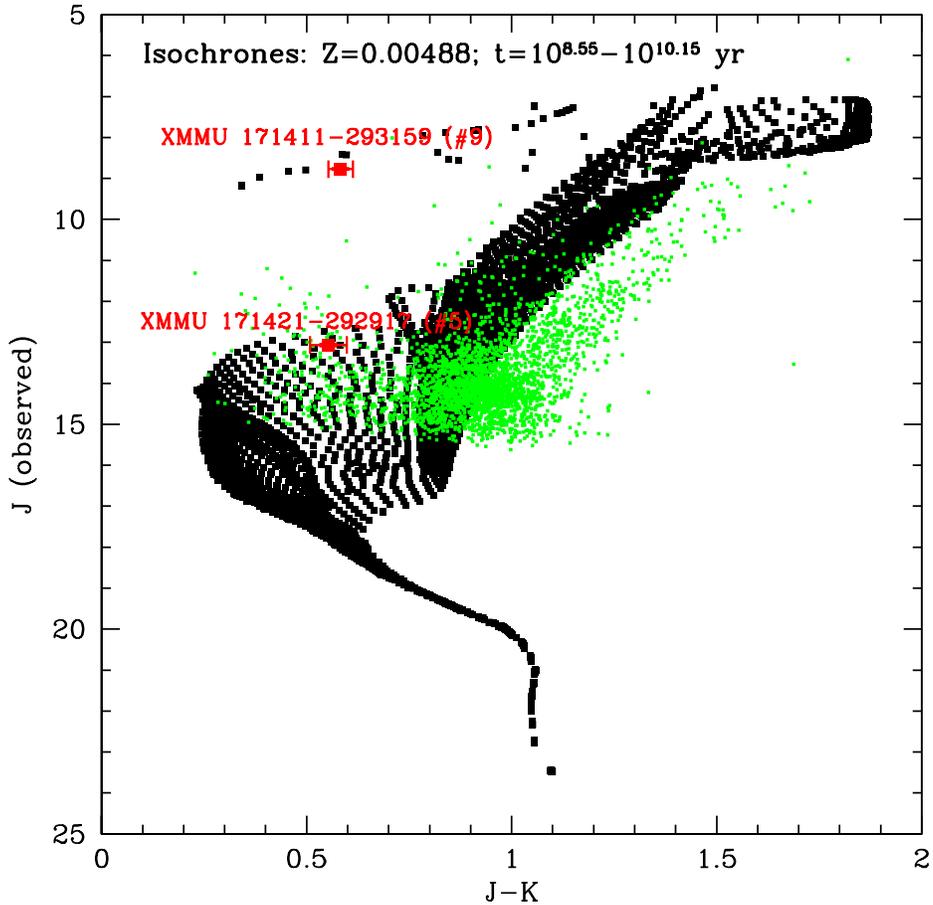,width=13.0cm,height=13.0cm,angle=0}~}
  \bigskip
  \caption[]{Color (J-K) Magnitude (K-band) diagram of the globular
    cluster NGC~6304 showing isochrones (small squares) for ages in
    the range $t = \ee{8.55}-\ee{10.15}\unit{yr}$ and $Z = 0.00488$
    \citep{marigo08}.  All 2MASS catalogued stars (small dots) within
    6.5 arcmin of the optical centre of the cluster are included,
    down to a magnitude of $m_{K}\sim15$.  The large squares with
    error bars are the IR counterparts to source \#9 and \#5, with
    $m_{J}=8.769$ and $m_{K}=13.063$, respectively. This diagram
    suggests that the counterpart for \sourcenine\ belongs to the
    post-asymptotic giant branch.  A spatial analysis of the stars
    that lie redward of the giant branch compared to those that lie
    blueward shows that there is an over-density of bluer stars
    near the GC core.  In other words, redder stars are less likely to
    belong to the GC and are likely to be contaminating
    field stars.  Thus, contamination by field stars can explain
    the poor matching between the theoretical isochrones and the
    actual data.
    \label{fig:JKJ}}
\end{figure}

\begin{figure}
  \centerline{~\psfig{file=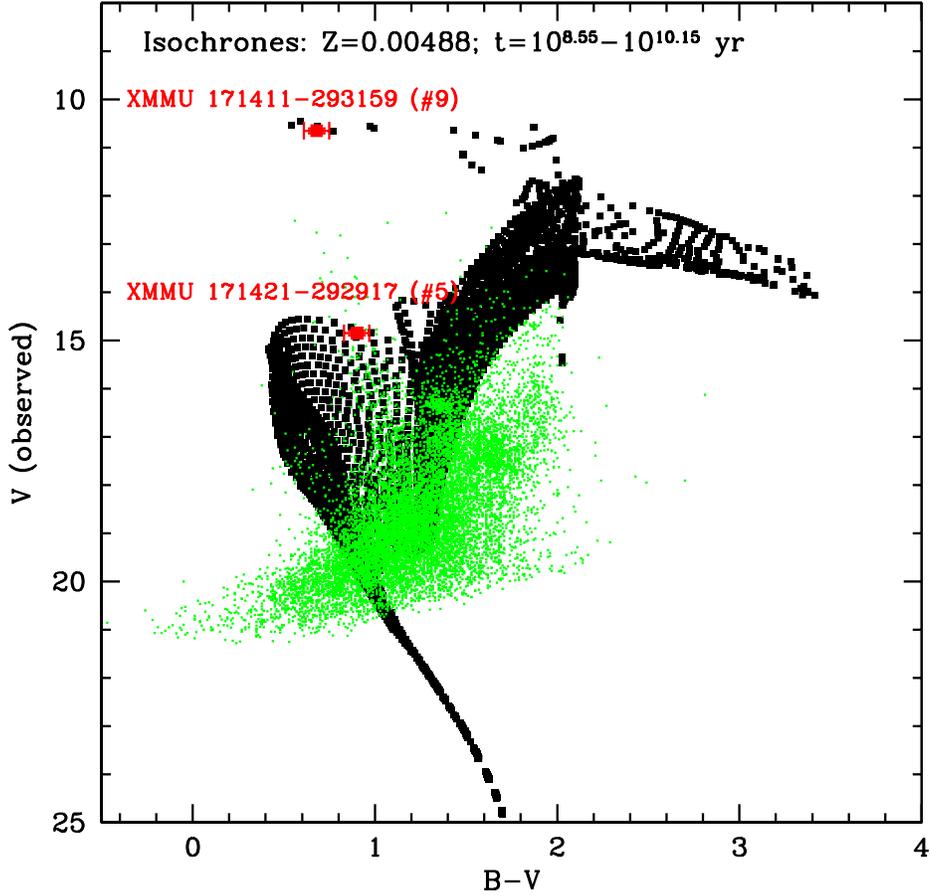,width=13.0cm,height=13.0cm,angle=0}~}
  \bigskip
  \caption[]{Color (B-V) Magnitude (V-band) diagram of the globular
    cluster NGC~6304 with the isochrones (small squares) for ages in
    the range $t = \ee{8.55}-\ee{10.15}\unit{yr}$ and $Z = 0.00488$
    \citep{marigo08}.  Also included are all the stars (small dots)
    within 6.5 arcmin of the optical centre (S. Ortolani, private
    communication, 2008 and \cite{ortolani00}) down to a limiting
    magnitude of $V\sim21$.  The large squares with error bars are the
    optical counterparts to source \#9 and \#5, with $V=10.65$ and
    $V=14.85$, respectively.  Similarly to Fig~\ref{fig:JKJ}, this
    suggests that the counterpart for \sourcenine\ is a post-AGB star
    at the distance of NGC~6304.  Also, field contamination could be
    responsible for the poor matching between the theoretical
    isochrones and the actual 2MASS data.
    \label{fig:VBV}}
\end{figure}

The color-magnitude diagrams (Figs.~\ref{fig:JKJ} \& \ref{fig:VBV})
containing theoretical isochrones for ages in the range $t =
\ee{8.55}-\ee{10.15}\unit{yr}$ and metallicity Z=0.00488
\citep{marigo08} show that the optical counterpart is consistent with
a post-asymptotic giant branch (post-AGB) star, both in $V$ and $J$
bands.  Even if the time scale of this stage is short, finding such an
object in a GC is not completely unexpected.  Based on
previous work, 16 planetary nebulae (PN) are expected to be found in
the GCs system of the galaxy \citep{jacoby97}. This corresponds to
$6.7\tee{-7}\unit{PN\, \perval{\Lsun}{-1}}$, using the total
luminosity of the GCs system, $\sim2.4\tee{7}\lsun$ \citep{secker92}.
Using the estimated mass of NGC~6304 \citep{gnedin02} \footnote{Online
data at http://www.astro.lsa.umich.edu/$\sim$ognedin/gc/vesc.dat} and
the average mass-to-light ratio of GCs, $M/L = 1.7$
\citep{caputo85}, 0.25 PNs are expected in this cluster.  Since the
lifetime of a PN, $\sim\ee{4}\unit{yr}$ \citep{jacoby97}, is
comparable to that of a post-AGB object, $\sim\ee{3}-\ee{4}\unit{yr}$
\citep{siodmiak08, blocker95}, finding one post-AGB star in NGC~6304
is not completely unexpected, and is consistent with the object
classification.

To obtain the bolometric luminosity of \TwoMassnine\ at the distance
of NGC~6304, the reddening $E(K-V)$ is first calculated using
$E(K-V)/E(B-V) = -2.744$ \citep{rieke85}, where $E(B-V)=0.53$ for
NGC~6304.  The intrinsic color is $(V-K)_{0} = (V-K) + E(K-V) = 2.38 -
1.45$.  Using bolometric corrections obtained from 2MASS photometry
\citep{masana06} as a function of $(V-K)_{0}$ and of the average
metallicity of the cluster $[m/H]=-0.56$ \citep{valenti07}, the
$K$-band bolometric correction is $BC_{K} = 0.892$. Using a $K$-band
extinction correction, $A_K=0.36\,E(B-V)$ \citep{fitzpatrick99},
$m_{\rm bol} = m_{K} - A_K + BC_{K} = 8.914$ is finally calculated to
be able to work out the bolometric luminosity.  Taking a zero absolute
bolometric magnitude $M_{\rm bol}=0$ star to correspond to a
luminosity of 2.97\tee{35}\cgslum\ \citep{harwit06} and the distance
modulus, the absolute bolometric magnitude is $M_{\rm bol}=-4.96$, for
a bolometric luminosity of 2.9\tee{37}\cgslum.

As a check on the value calculated for the bolometric luminosity, a
less precise method of estimation is used (since it does not take into
account effects of realistic atmospheres, such as limb darkening), by
fitting a blackbody curve to the unabsorbed photometry in $BVJHK$
bands.  For a blackbody source at the distance of NGC~6304, this
resulted in a best fit ($\chi^2/$dof = 5.8, for 3 dof) of $R=46\ppm1.0
\rsun$, $T_{\rm eff}=6620\ppm140\unit{K}$, for a total luminosity of
$L=4\pi R^2 \sigma T_{\rm eff}^4=1.4\tee{37}\cgslum$; this is a factor
of $\times2.1$ below that derived through the more precise method,
which may be indicative of the systematic uncertainties in
approximating the total luminosity using a blackbody fit to the
$BVHJK$ data, but which is regarded as consistent with the more
precise $BC_{K}$ calculation method.

\subsubsection{Examining the Hypothesis that \sourcenine\ is a Foreground Star} 
The hypothesis that \TwoMassnine\ is a field dwarf star is
investigated.  First, this star (TYC 6824-713) is listed in the
Tycho-2 catalog (issued from a re-analysis of the Hipparcos data)
providing proper motion, but no parallax information \citep{hog00}.
The proper motion observed for this star is --6.6\ppm2.8
milli-arcseconds per year, marginally consistent (at $2.4\sigma$) with
a zero proper motion, as would be expected for a cluster member.  This
does not resolve the possible cluster membership versus the foreground
star hypothesis.

If reddening dust is distributed uniformly between the observer and
NGC~6304, the source, assuming it is a main-sequence star, is most
consistent with an $M_V=4.4$ object at a distance 170\pc, with
$E(B-V)=0.03$, based on the absolute magnitude $M_V$ and color $B-V$
values of stars in the Hyades \citep{debruijne01}.

For main-sequence stars, the typical \xray\ to bolometric flux ratio
seems to reach an upper limit at $\ee{-3}$ \citep{vilhu87}. For the
association \sourcenine\slash\TwoMassnine, the \xray\ to bolometric
flux ratio is of the order of $\sim\ee{-5}$, therefore consistent with
a main-sequence foreground star.

\begin{figure}
  \centerline{~\psfig{file=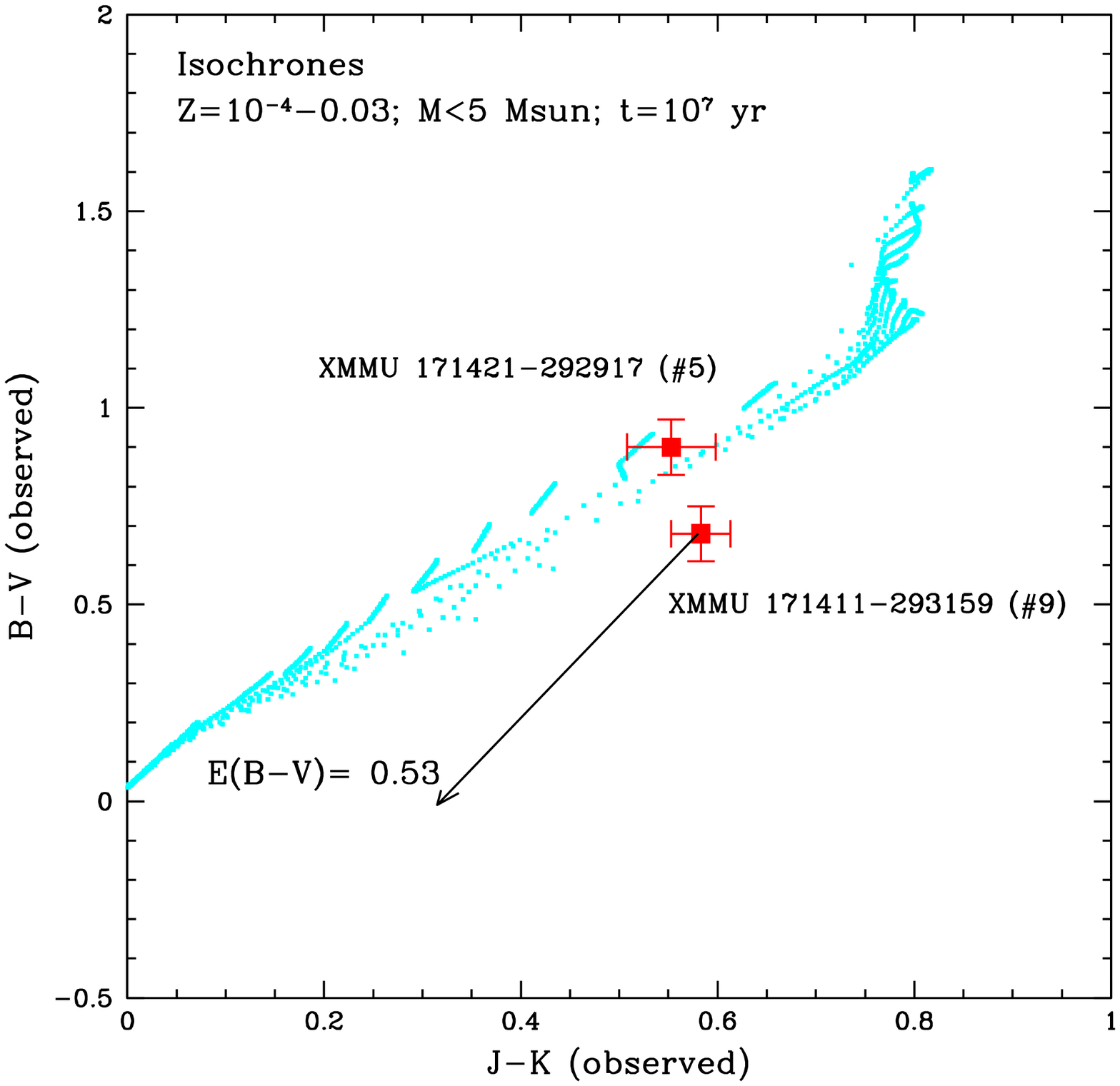,width=13.0cm,height=13.0cm,angle=0}~}
  \bigskip
  \caption[]{Color ($J-K$) Color ($B-V$) diagram showing isochrones of
    low-mass ($M<5\msun$) stars at $t=\ee{7}\unit{yr}$ (small points)
    for a range of metallicities ($Z=0.0001-0.03$, $\Delta Z=0.0005$).
    The large points with error bars are corresponding to \sourcefive\
    (\#5) and \sourcenine\ (\#9).  In considering source \#9 as a
    foreground coronally active star, it lies marginally off
    the main sequence (between 2-3$\sigma$) at zero reddening
    $(E(B-V)=0)$.  The discrepancy worsens if source \#9 is at greater
    reddening; the vector beginning at source \#9 terminates at the
    intrinsic color \TwoMassnine\ would be at $E(B-V)=0.53$, that of NGC
    6304.  This does not support the hypothesis that \sourcenine\ is a
    typical coronally active star, in the foreground of NGC~6304.
    \label{fig:bvjkccd}}
\end{figure}

In a $B-V$ vs.\ $J-K$ color-color diagram (Fig.~\ref{fig:bvjkccd}),
\sourcenine\ lies marginally off the main-sequence (by between 2 and 3
$\sigma$) for a wide range of metallicities ($Z=0.0001-0.03$), lying
closest to stars in the mass range 0.7--0.9\msun, depending on
metallicity, according to theoretical isochrones \citep{marigo08}.
This marginal result from the color-color diagram does not
definitively exclude the possibility that the star is in the
foreground.

The X-ray emission from dMe stars is spectrally comparable to those of
RS CVn \citep{singh96}.  Typically, RS CVn exhibit a two-temperature
Raymond-Smith plasma model \citep{dempsey93b}.  Since the number of
counts are not sufficient for such a fit with all parameters left
free, the spectrum of \sourcenine\ is fitted twice with a
two-temperature Raymond-Smith plasma model keeping the high
temperature component fixed at the lowest ($kT_{2}=0.93\keV$) and
highest value ($kT_{2}=3.45\keV$) from the sample of RS CVn
\citep{dempsey93b}. The metallicities are also held fixed at the solar
value, $Z = 0.0177$.  The normalization of the Raymond-Smith model in
{\tt XSPEC} model provides the emission measure (${\rm EM}=\int
n_{e}n_{H} dV$), a distance dependent value.  The best fit parameters
for those two statistically acceptable fits are
$\nhtt=1.3\ud{0.6}{0.4}$, $kT_{1}=0.11\ud{0.04}{0.04}\keV$, ${\rm
EM_{1}}=1.9\ud{11.2}{1.8} \tee{60} \percmcube$ and ${\rm EM_{2}}/{\rm
EM_{1}}=1\ud{5}{1} \tee{-5}$ for the first fit corresponding to
$kT_{2}=0.93\keV$ and $\nhtt=0.14\ud{0.24}{0.10}$,
$kT_{1}=0.84\ud{0.18}{0.25} \keV$, ${\rm EM_{1}}=1.3\ud{0.4}{0.4}
\tee{56} \percmcube$ and ${\rm EM_{2}}/{\rm EM_{1}} \leq 0.55$ for
$kT_{2}=3.45\keV$.

The value of ${\rm EM_{1}}$ lies between 3 and 7 orders of magnitude
above the typical values (${\rm EM_{1}} = 0.1-3\tee{53}\percmcube$),
for the fits performed with the maximum and minimum $kT_{2}$
respectively, assuming \TwoMassnine\ lies in NGC~6304.  Thus, an
object with a typical ${\rm EM}_1$ would have to be at a distance less
than $\sim330\pc$.

The ratio of the two emission measures ${\rm EM_{2}}/{\rm EM_{1}}$,
however, is a distance independent value, that is in all cases $\geq
1$ for coronally active stars \citep{dempsey93b}.  In both cases, the
derived 90 per cent confidence upper limits given before lie below
values observed from coronally active systems.  Therefore, the
observed X-ray spectrum is inconsistent with the typical
two-temperature plasma of a coronally active star, independently of the
distance of the IR counterpart.

In conclusion, the optical/IR colors and proper motion of the
counterpart are in marginal agreement with those of a main sequence
object located in the foreground of NGC~6304.  However, the X-ray
spectrum of \sourcenine\ is inconsistent with those of typical
coronally active stars.  This was shown using a distance independent
value.  In addition, the possible optical variability of the star (in
the $B$-band, see Sec.~\ref{sec:searchIRC}) presents further support
that the star is not merely a typical dwarf main sequence star.  At
the same time, the X-ray spectrum is consistent with a H atmosphere
qLMXB at the distance of NGC~6304; and the V vs. B-V and J vs. J-K
CMDs support interpretation of the 2MASS counterpart as a post-AGB
star in NGC~6304, which is evolutionarily consistent with the X-ray
spectrum of a qLMXB.  Without excluding classification as a foreground
star, we identify \sourcenine\ as a candidate qLMXB in NGC~6304;
however, an optical/IR spectrum of the 2MASS counterpart should be
obtained to confirm that the IR source is an actual post-AGB star in
NGC~6304, and not a marginally spectrally unusual, optically variable
main-sequence star (0.7--0.9\msun) at a distance of $\sim$few 100\pc.

\subsection{\sourcefive\ - Source \#5}
\label{sec:sourcefive}

\begin{figure}
  \centerline{~\psfig{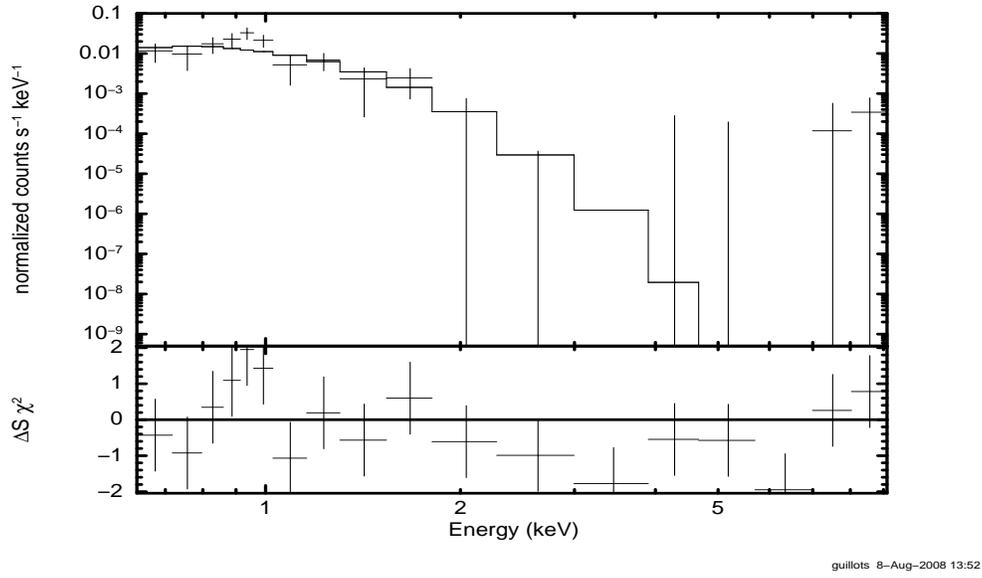}~}
  \bigskip
  \caption[]{Folded model Spectrum of source \#5 (\sourcefive).  The
    solid line is the best fit (tabulated) model of a neutron star
    atmosphere ($\nhtt=0.266$ held fixed).
    \label{fig:Obj5-folded}}
\end{figure}

The NS H-atmosphere tabulated model applies for this low
signal-to-noise source.  The fit is statistically acceptable
(\Chisq{1.09}{16}{0.36}) and the parameters are consistent with
expected values ($\kteff=70 \ud{28}{20}\eV$ and
$\rinfty=23\ud{69}{10}\km$ with $\nhtt=0.266$; See
Fig~\ref{fig:Obj5-folded}).  When leaving \nh\ free, the fit results
become \Chisq{1.12}{15}{0.33} with $\nhtt=0.46\ud{0.42}{0.68}$,
$\kteff=49\ud{77}{32}\eV$ and $\rinfty\geq8.9\km$ (the upper bound
error estimate failed to converge), i.e. consistent with the previous
fit when \nh\ was fixed.

The {\tt nsagrav} model is then used to confirm the first fit with the
NS H-atmosphere model.  Even though acceptable
(\Chisq{1.23}{16}{0.23}), the fit of this low signal-to-noise spectrum
does not allow for representative error estimates, as explained in
\S\ref{sec:spectra}.  The other two {\tt XSPEC} neutron star
atmosphere models also have the same issue. Also, the simultaneous
fitting of the pn, MOS1 and MOS2 spectra did not provide an
improvement of the uncertainties.

Using the online 2MASS-PSC catalogue, a possible counterpart,
\TwoMassfive\ ($m_{J}= 13.063(29)$, $m_{H}= 12.645(33)$ and $m_{K}=
12.510(35)$), is identified at an angular distance of 0.45 arcsec.
Using the same calculations as for source \#9, the probability of
finding a counterpart as bright or brighter, as close or closer is
0.084 per cent when 179 stars with magnitude $m_{J} \leq 13.063$ are
found in an annulus of inner radius 2 arcmin and outer radius 4 arcmin
around the optical centre of NGC~6304\footnote{Prior to the
astrometric correction using \sourcenine\ and \TwoMassnine, the
angular distance between the \xray\ and 2MASS sources was 1.59 arcsec.
The corresponding association confidence would have been, in that
case, 98.95 per cent.}, implying an association with 99.916 per cent
confidence.

Using the procedure described in \S~\ref{sec:searchIRC}, the
bolometric magnitude and luminosity of the IR counterpart are
calculated from the V-band magnitude $V=14.85(5)$, the B-band
magnitude $B=14.75(9)$ and the color $B-V=0.90(7)$ (S. Ortolani,
private communication, 2008).  The values found are $m_{\rm bol} =
m_{K} - A_{K} + BC_{K} = 12.82$, implying an absolute bolometric
magnitude $M_{\rm bol} = -1.07$.  Therefore, the bolometric luminosity
is $L_{\rm bol} = 7.96\tee{35}\cgslum$, when the bolometric luminosity
for the zero absolute bolometric magnitude is 2.97\tee{35}\cgslum. As
a check on the value calculated for the bolometric luminosity, a
blackbody curve is fitted to the unabsorbed photometry in $BVJHK$
bands, assuming the source at the distance of NGC~6304.  This resulted
in a best fit ($\chi^2/$dof = 3.98, for 3 dof) of
$R=5.87\ppm0.14\rsun$, $T_{\rm eff}=7180\ppm150\unit{K}$, for a total
luminosity of $L=4\pi R^2 \sigma T_{\rm eff}^4=3.15\tee{35}\cgslum$;
this is a factor $\times2.5$ below that calculated using the $BC_K$
calculation above.

If the \xray\ emission is coming from the star \TwoMassfive, the
\xray\ to optical flux ratio for this giant star is therefore
$F_{X}/F_{\rm bol} = \ee {-4.05}$.  From the ROSAT All Sky Survey
catalogue of bright late-type giant and supergiant \citep{hunsch98},
the mean calculated \xray\ to optical flux ratio for all 450 stars
observed in that catalogue is $F_{X}/F_{\rm bol}=10^{-5.6\pm{0.6}}
\,(1\sigma)$.  The 0.1--2.4\keV\ range was used to estimate the \xray\
flux.  The 2MASS counterpart and the \xray\ source \#5 are consistent
with being a giant star $(2.6\sigma)$.  However, if X-rays from
\sourcefive\ were due to a typical giant star, then similar \xray\
flux from other giant stars in this GC would be expected; yet no other
giant star on the outskirts of this GC exhibit similar \xray\ fluxes.
Also, using the same comparison as in Sec~\ref{sec:searchIRC}, the
\xray\ to bolometric flux ratio, $F_{X}/F_{\rm bol} =9\tee{-5}$ is
consistent with a main-sequence foreground star, for which the \xray\
to bolometric flux ratio saturates at $\ee{-3}$ \citep{vilhu87}.

An analysis similar to the one performed for \sourcenine\
(\S~\ref{sec:sourcenine}) excludes the possibility that the system is
coronally active.  The results of the spectral fits with a
2-temperature Raymond-smith (RS) plasma are as follows.  For
$kT_{2}=0.93\keV$ (\Chisq{1.25}{14}{0.23}), $\nhtt=1.15
\ud{1.27}{0.36}$, $kT_{1}=0.11 \ud{0.22}{0.06} \keV$, ${\rm
EM}_{1}=2.4 \ud{12.9}{2.1}\tee{59} \percmcube$ and ${\rm EM}_{2}
\leq3.9\tee{55} \percmcube$.  In the case of $kT_{2}=3.45 \keV$
(\Chisq{1.25}{14}{0.23}), $\nhtt=1.12\ud{1.12}{0.53}$, $kT_{1}=0.11
\ud{0.19}{0.06} \keV$, ${\rm EM}_{1}=2.1 \ud{18.5}{1.5}\tee{59}
\percmcube$ and ${\rm EM}_{2}\leq5.5\tee{54} \percmcube$.  Therefore,
the distance independent value ${\rm EM_{2}}/{\rm EM_{1}}$ is
$\approxlt \ee{-3}$ in both cases and lower than the value ${\rm
EM_{2}}/{\rm EM_{1}} \geq 1$ observed for known coronally active stars
\citep{dempsey93b}.  Thus, the X-ray spectrum of this source is
inconsistent with that of typical coronally active stars.  In
addition, the best fit galactic absorption \nh values, $\nhtt=1.15
\ud{1.27}{0.36}$ or $\nhtt=1.12\ud{1.12}{0.53}$, for the two cases of
the two-tempeture RS plasma, are significantly larger than the value
in the direction of NGC~6304 ($\nhtt=0.266$).  Therefore, assuming an
uniform distribution of absorbing material, the best fit value of \nh\
provides further support that \sourcefive\ is unlikely to be a
coronally active star in the foreground of the GC.

We therefore conclude that that \sourcefive\ is not a coronally active
foreground star.  Finally, since the \xray\ source is spectrally
consistent with a qLMXB at a distance of 6\kpc, we identify the X-ray
source as a candidate qLMXB in NGC~6304 with an identified IR
companion in NGC~6304.

To estimate the accretion rate, the series of relations in
\cite{verbuntBOOK} (X-Ray binaries, section 11.3.3) are used.  From
the radius of the companion star ($R_{2}=9.7\rsun$), calculated with
the temperature $T_{\rm eff}=6915\unit{K}$ (which is worked out from
\cite{masana06}) and the luminosity $L_{\rm bol} =7.96\tee{35}
\cgslum$, the companion core mass: $M_{\rm c}=0.27\msun$ ($Z=0.02$)
and $M_{\rm c}=0.31\msun$ ($Z=0.0001$) are estimated, where
$y\equiv\ln \left( M_{\rm c}/0.25\msun \right)$ with $\ln \left(
R_{2}/\Rsun \right)$ and $\ln \left( L_{2}/\Lsun \right)$ being third
order polynomials of $y$ where the polynomial coefficients are
theoretical fits given in \cite{verbuntBOOK}.  From $M_{\rm c}$ and
$\dot{M_{\rm c}} = 1.37\tee{-11}\left( L/\Lsun \right)
\unit{\Msun\peryear}$, $\dot{R_{2}}/R_{2}$ is estimated giving the
evolution time-scale of the companion: $\tau \approx
2\tee{7}\unit{yr}$ ($Z=0.02$) and $\tau \approx 2.3\tee{7}\unit{yr}$
($Z=0.0001$).  Finally, the following relation:
\begin{equation}
  \frac{\dot{R_{2}}}{R_{2}} = -2\,\frac{\dot{M_{2}}}{M_{2}}\left(
  \frac{5}{6} - \frac{M_{2}}{M_{1}}\right)
\end{equation}
\noindent is used to calculate the accretion rate $\dot{M_{2}}$ from
the companion on to the neutron star, assuming $M_{2}=0.8\msun$.  We
found that $\dot{M_{2}}=-4.4\tee{-9}\unit{\msun\peryear}$ ($Z=0.02$)
and $\dot{M_{2}}=-3.8\tee{-9}\unit{\msun\peryear}$ ($Z=0.0001$).  With
this value of $\dot{M_{2}}$ (for $Z=0.02$), the quiescent luminosity
in the DCH model is estimated using models that include core neutrino
emission \citep{yakovlev03}.  The thermal \xray\ luminosity, $L_{\rm
thermal}=1.1\tee{33}\cgslum$, is roughly consistent with thermal
emission from other transients with high time-average mass transfer
rates \citep{heinke07}.

\section{Discussion}
\label{sec:discuss}

\subsection{Comparison with {\em ROSAT} Observations} 
\label{sec:rosat}

NGC~6304 was observed once previously, using \rosat/HRI \citep[R94
hereafter]{rappaport94}, in which four \xray\ sources were discovered.
Three are spatially coincident with three \xray\ sources detected in
the present observations (in fact, the three correspond to the
candidate qLMXBs).  The fourth \xray\ source in R94, however, appears
to have faded significantly; the present observation detects no \xray\
source consistent in position with \rosat\ source D.  The following
analysis compares the position and fluxes of the for ROSAT/HRI
sources.

The positions of the four sources in R94 have been examined and
compared to those of X-ray sources in the present work.  No positional
uncertainties are given from analysis of R94.  Using the archived
\rosat/HRI data at {\it HEASARC}, the positional uncertainties
required for the comparison with the present \xmm\ data were obtained.
The source positions are consistent between the two different
observations: source A, B and C in R94 corresponding to sources \#4,
\#5 and \#9 in this work.  The boresight systematic uncertainty (6
arcsec) have been taken into account.  For source D, no corresponding
source is detected within 1 arcmin of the position for source D given
by R94.  A total of 304 counts are found in a 25 arcsec area around
the position given by R94, while an average of 294.8\ppm2.2 are due to
background using a nearby off-source area.  This leaves 9\ppm18 counts
due to a possible X-ray source at this position, consistent with no
source detection.

There is a $3\sigma$ upper limit on the flux from source D of
$F_{X}\leq2.4\tee{-14}\cgsflux$ (0.5--2.4\keV), assuming a thermal
bremsstrahlung spectrum with a temperature $kT=3\keV$ (R94).  By
inspection of the MOS2 data, there is no evidence of any \xray\ source
in the vicinity of the position of source D.

Table \ref{tab:CompFluxB} compares the predicted number of source
counts which should be detected with \rosat/HRI during the 5030 sec
observation of R94, assuming the observed XMM/pn-med countrates and
the XMM/pn source spectra, including both countrate and spectral
uncertainties.  For sources \#4, \#5, and \#9, the expected number of
counts (20.7\ppm1.5, 3.9\ppm0.9, 11.4\ppm1.0, respectively) are
consistent, within Poisson uncertainties, with the number of source
counts observed by R94 (15, 7, and 17 respectively).

For source D, which is not detected, the 3$\sigma$ upper limit on the
number of counts due to an X-ray source at this position ($<$63
counts, in 11037 sec of integration), implies that $<$1.8 counts
should have been detected with \rosat/HRI, whereas the observed number
of counts with \rosat/HRI corresponds to 14 counts at the HRI field
centre\footnote{The number of counts given by R94 are corrected for
scattering and vignetting by a factor $>$1, but which is not given by
R94.  Therefore while R94 gives the number of counts which would have
been detected from this source if it had been located at the centre of
the FOV, the number of counts detected was not noted, and so the
uncertainty in the number of counts has not been determined.}.  In an
absence of an error analysis on the number of detected source counts
from R94, it is not possible to correctly estimate the amount of
fading; however, taking the detected number of counts at face value
implies that source D faded in luminosity by a factor of $\sim$10.

In conclusion, the three qLMXBs are consistent with having the same
luminosity between the \rosat/HRI observation and the XMM/pn
observation; and source D appears to have faded by a factor of
$\sim$10.

\subsection{Expected number of quiescent Neutron Stars in Globular 
Clusters}
\label{sec:number}

The number of \xray\ binaries in a GC can be predicted from its
internal properties.  The encounter rate for a GC is predicted by
$\Gamma\propto\rho_{0}^{2}/v$ where $\rho_{0}$ is the central
luminosity density and $v$ is the velocity dispersion
\citep{verbunt02}.  For a virialized cluster, this relation simplifies
to $\Gamma\propto\rho_{0}^{1.5}r_{c}^{2}$ where $r_{c}$ is the core
radius (in physical units, not angular distance).  The number of
quiescent NSs found in NGC~6304 is compared to the expected number of
such objects from previous works \citep{gendre03b,heinke03b}.

In the first one (Table 2 and Figure 3 in \cite{gendre03b}), the
authors found a linear relation between the number of quiescent NSs
and the encounter rate: $N_{qNS}\sim0.04\times\Gamma+0.2$.  Using the
same normalisation ($\Gamma_{\rm NGC~6440}=100$), the encounter rate
of NGC~6304 is $\Gamma_{\rm NGC~6304}=6.4$.  The predicted number of
qLMXBs is therefore about 0.46.  Assuming that three qLMXBs were
found, this number seems slightly in excess of the prediction but the
probability of finding such objects when the average expected number
is 0.46 is 3.2 per cent , assuming Poisson statistics.  This
probability is not small enough to affirm that there is an
inconsistency between the expected and the actual number.

The second paper (Table 1 in \cite{heinke03b}) compared the encounter
rate (normalised to the galactic value) to the number of accreting
NS systems.  The linear relation obtained from the data collected in
this previous work is simply $N=0.993\times\Gamma-0.046$, for which
the expected number of accreting NSs in NGC~6304 is 0.38.  Again,
three qLMXBs is a number that seems larger than the expectation, but
certainly not inconsistent.

Finally, the encounter rate can be compared to the number of \xray\
sources in the GC \citep{pooley03}.  The expected number
of \xray\ sources for an encounter rate $\Gamma_{NGC~6304}=45$ is
about 9.5, comparable with the seventeen sources detected in the FOV.
The following subsection discuss the number of background sources that
are expected to be detected in the field of view of the observation.

\subsection{Background \xray\ sources expected in the field of NGC~6304}

To estimate the number of expected Active Galactic Nuclei (AGN)
detected during the observation of NGC~6304, the field of the pn
camera is divided into 3 concentric annulli centred at the centre of
the exposure map (with radii between 0 and 5, 5 and 10, and between 10
and 13 arcmin).  The limiting flux in each annulus is estimated using
the lowest count rate detected (4.3 counts per sec) and the tool {\it
webPIMMS}, to obtain the absorbed flux a power-law of photon index 2.
The limiting fluxes are $1.8\tee{-14}\cgsflux$,
$1.2\tee{-14}\cgsflux$, and $0.9\tee{-14}\cgsflux$, for the three
annulli respectively. Using the number-flux (LogN-LogS) distribution
function for X-ray AGN \cite[H93 hereafter]{hasinger93}, the number of
expected background \xray\ sources is 0.87\ppm0.1, 5.22\ppm0.28 and
10.7\ppm0.75 in the three annulli, respectively, corresponding to a
total of 17\ppm1 over the whole field.  This result is consistent with
the number of sources detected in the field, and with the tentative
conclusion claiming that 3 qLMXBs are detected in the GC NGC~6304.
However, the error bars only takes into account the uncertainty in the
LogN-LogS relation and not in the limiting fluxes and are therefore
likely to be underestimated.

\section{Summary and Conclusions}
\label{sec:conclude}

The globular cluster NGC~6304 hosts three candidate qLMXBs.  The
neutron star radii and temperatures obtained for the three candidates
are consistent with typical values for neutron stars and in accordance
with radii of other known field and GCs qLMXBs.  No variability is
measured from any of the candidates over the time scale of the
observation, with very weak limits; the large X-ray variability in the
\xmm/pn background precludes a more detailed variability analysis.  A
comparison with \rosat\ observations 14 yr prior (R94), finds the
fluxes of the candidates are consistent with those of the earlier
observation (R94).  The absence of a detailed error analysis in the
fluxes from the published \rosat\ analysis precludes placing
statistical limit on the amount of variability; however, as the number
of detected counts in the \rosat\ observations was small ($<$20), the
magnitude of intensity variability is limited to be less than a factor
of $\sim$few on the 14-yr time-scale.

\sourcefour\ is located in the core (0.79\unit{r_c}) of the GC.  Its
spectrum is acceptably fitted with a NS Hydrogen atmosphere model with
$\kteff=122 \ud{31}{45}\eV$ and $\rinfty=11.6\ud{6.3}{4.6}\km$,
combined with a power-law component of photon index $\alpha=
1.2\ud{0.7}{0.8}$.  The power-law component dominates the spectrum at
photon energies above 2\keV, and represents 49 per cent of the
observed flux (0.5--10 keV).  This contribution, although its source
is not perfectly understood, is similar to that found in some other
qLMXBs, and is comparable in the fraction of the flux for which it
accounts to the field qLMXB Cen~X-4
\citep{asai96b,rutledge01,menou01,campana04}. While the high stellar
density in the core of NGC~6304 precludes identification of a 2MASS
counterpart for \sourcefour, a visual overlap with a 2MASS image of
the GC shows a possible association with an uncatalogued -- and likely
unresolved -- star.  The association will require further
investigation with higher resolution optical/IR imaging.

\sourcefive, a second candidate, has a low signal-to-noise and
correspondingly larger error bars for the X-ray spectral parameters
used to identify it as a qLMXB candidate: the NS H-atmosphere
temperature is $70 \ud{28}{20}\eV$, and $\rinfty=23\ud{69}{10}\km$.  A
faint 2MASS counterpart ($m_{J}=13.063$) is identified (with 99.916
per cent confidence), and the \xray\ emission is unlikely to emerge
from the giant star itself.

\sourcenine, a third candidate, is a bright \xray\ source consistent
with a neutron star H atmosphere spectral model at the distance of the
GC.  Its measured properties are $\kteff=115 \ud{21}{16}\eV$ and
$\rinfty = 10.7\ud{6.3}{3.1}\km$.  The 2MASS IR counterpart (with
99.988 per cent confidence) appears to belong to the post-asymptotic
giant branch; this is the first qLMXB with this type of companion.
Other explanations for the source are investigated, concluding that a
coronally active field star is not a possible explanation, due to the
unusual optical/IR colors and unusual X-ray spectrum.  Optical/IR
spectroscopy of \TwoMassnine\ can definitively determine whether the
star is a foreground main sequence star, or an evolved post-AGB star
at the distance of NGC~6304.

A faint unidentified \xray\ source showing an unusual excess of high
energy photons resulting in a photon index of $\alpha =
-2.0\ud{1.2}{2.2}$.  The nature of this source is unclear. There is no
evidence the X-ray source was variable during the observation.  No
off-band counterparts in the 2MASS or DSS catalogues are found within
$3\sigma$ of the X-ray position.  Deep observations in the off-bands,
as well as repeated detection and study in the X-ray, can inform
interpretation of this source.

The qLMXB candidates in NGC~6304 are astronomically interesting.
First, the number of candidates is consistent with predictions from
the calculated encounter rate, supporting previous observational
analyses which characterised the qLMXB rate vs. encounter rate in GCs
\citep{pooley03,heinke03b,gendre03b}.  This indicates that the qLMXB
rate vs. encounter rate is not merely a parameterisation resulting
from observations, but can predict the number of qLMXBs in unobserved
GCs.

NGC~6304 contains one of the most luminous known qLMXBs.  \sourcefour,
with a intrinsic luminosity $L_{X}=\ee{33}\cgslum$ (0.5--10\keV), is
the fourth most luminous GC qLMXB after 47 Tuc X7
($L_{X}=1.5\tee{33}\cgslum$ (0.5--10\keV)), 47 Tuc X5
($L_{X}=1.4\tee{33}\cgslum$ (0.5--2.5\keV)) and the qLMXB in M28
($L_{X}=1.2\tee{33}\cgslum$ (0.5--10\keV)).  These represent the upper
limit of the expected luminosity for LMXB in quiescence.  In addition,
\sourcefour, with its high flux ($F_{X}=2.3\tee{-13}\cgsflux$, after
47 Tuc X7, 47 Tuc X5 and the qLMXB in M28 with fluxes 5.3, 4.3 and
$3.3\ud{1.9}{1.1}$ $\tee{-13}\cgsflux$, respectively - see
Table~\ref{tab:NS}), is a good candidate to constrain the equation of
state of dense matter.

NGC~6304 hosts two candidate qLMXBs lying at large distance from the
hosting core, but still within the tidal radius; \sourcenine\ and
\sourcefive\ are located at $\sim32r_{c}$ and $\sim15r_{c}$,
respectively.  Previously, all GC qLMXBs were located within $<7 r_c$
of the optical centre of their GC (see Table~\ref{tab:Rcore}), the
most distant belonging to U24 in NGC~6397, located $6.8r_{c}$ from the
centre; noting, however, that NGC~6397 is a core collapse GC with
$r_{c}=0.05\arcmin$.  To sum up, the qLMXB candidates in NGC~6304 are
of much greater distance from the GC centre than has been observed
previously, and may require revisiting theory of qLMXB formation and
binary evolution in GC.

Deeper X-ray exposures are required to confirm the classification of
the candidates qLMXB found in NGC~6304.  Specifically, it is necessary
to constrain \rinfty\ to an uncertainty of few percent (as already
performed for confirmed qLMXBs in $\omega$ Cen and 47 Tuc, see Table
\ref{tab:NS}) rather than few tens of percent here.  For \sourcefour,
better angular resolution will confirm the isolated nature of the
candidate qLMXB, or may resolve multiple \xray\ sources in the core.
New deep observations of \sourcefive\ and \sourcenine\ will produce a
localisation unbiased by bad detector columns.  At optical
wavelengths, observations of the aforementioned candidates can
determine unequivocally the evolutionary state of the counterparts (a
giant for \TwoMassfive, and a post-AGB star for \TwoMassnine), as well
as measure their binary orbital periods through ellipsoidal
variations.  The spatial isolation and brightness of the IR
counterparts implies this work can be done easily using ground based,
moderate sized telescopes.  Optical spectroscopic observations can
confirm that the counterparts have radial velocities consistent with
that of NGC~6304.  Finally, high spatial resolution observations of
the GC core in the optical/IR bands can pinpoint a
possible counterpart for \sourcefour\ and confirm the qLMXB nature of
this \xray\ source.

\acknowledgements

Robert E. Rutledge is supported by an NSERC Discovery grant. Edward
F. Brown and Vyacheslav E. Zavlin acknowledge support from NASA under
award No.~NNX06AH79G.  The authors are particularly thankful to
Prof. Sergio Ortolani for providing the V-band magnitude data of stars
in the GC, especially for the two aforementioned
counterparts, to calculate the bolometric luminosities and produce the
$V$ vs $(B-V)$ diagram.  Finally, the authors would like to thank the
referee for fruitful remarks.

\clearpage


\bibliographystyle{apj_8}
\bibliography{complete}

\clearpage

\clearpage

\clearpage

\clearpage

\clearpage

\clearpage



\begin{deluxetable}{cccccc}
  \tablewidth{0pt} 
  \tablecaption{\xray\ Catalogue of sources in the field of NGC~6304 (Astrometry)} 
  \tabletypesize{\scriptsize}
  \tablehead{ 
    \colhead{Object Name} & \colhead{R.A.} & \colhead{Decl.} & 
    \colhead{$\delta_{\rm R.A.}\backslash \delta_{\rm Decl.}$\tablenotemark{a} } & 
    \colhead{$\Delta/r_c$\tablenotemark{b}} & \colhead{ID} \\
    \colhead{} & \colhead{(J2000)} & \colhead{(J2000)} & 
    \colhead{(arcsec)} & \colhead{} & \colhead{} \\ } 
  \startdata
  XMMU~171403$-$292438 & 258.51323 & -29.41030 & $\pm$0.5$\backslash$0.5 & 33.4 & 10 \\ 
  \sourcenine          & 258.54803 & -29.53317 & $\pm$0.5$\backslash$0.4 & 31.8 & 9 \\ 
  \sourceeight         & 258.55432 & -29.57151 & $\pm$1.0$\backslash$0.8 & 38.6 & 8 \\ 
  XMMU~171415$-$292232 & 258.56320 & -29.37568 & $\pm$0.9$\backslash$0.8 & 31.9 & 7 \\ 
  XMMU~171417$-$293222 & 258.57100 & -29.53968 & $\pm$0.9$\backslash$1.0 & 28.5 & 6 \\ 
  \sourcefive          & 258.58731 & -29.48801 & $\pm$0.8$\backslash$1.0 & 15.2 & 5 \\ 
  \sourcefour          & 258.63636 & -29.46310 & $\pm$0.7$\backslash$0.6 & 0.79 & 4 \\ 
  XMMU~171449$-$292001 & 258.70731 & -29.33378 & $\pm$1.0$\backslash$0.9 & 42.3 & 3 \\ 
  XMMU~171451$-$293653 & 258.71296 & -29.61490 & $\pm$1.0$\backslash$0.9 & 49.1 & 2 \\ 
  XMMU~171453$-$292436 & 258.72141 & -29.41001 & $\pm$0.7$\backslash$0.6 & 29.2 & 1 \\
  XMMU~171516$-$292224 & 258.81800 & -29.37320 & $\pm$0.5$\backslash$0.5 & 52.2 & 0\\
  \enddata
  \label{tab:all1}
  \tablecomments{The sources are sorted by increasing R.A.  Both
    coordinates (R.A. and Decl.) are given in decimal degrees, after
    the astrometric correction.}  
  \tablenotetext{a}{Statistical
    uncertainty from the source detection in seconds of arc.}
  \tablenotetext{b}{distance from GC centre in $r_c$ units -
    $r_c$=0.21\arcmin\ (Harris 96).}
\end{deluxetable}


\begin{deluxetable}{ccccccc}
  \tabletypesize{\scriptsize}
  \tablewidth{0pt}
  \tablecaption{\xray\ Catalogue of sources in the field of NGC~6304 detected with the MOS camera only}
  \tablehead{
    \colhead{Object Name} & \colhead{R.A.} & \colhead{Decl.} & 
    \colhead{$\delta_{\rm R.A.}\backslash \delta_{\rm Decl.}$\tablenotemark{a} } & 
    \colhead{S/N}  & \colhead{Rate} & \colhead{ID}\\
    \colhead{} & \colhead{(J2000)} & \colhead{(J2000)} & \colhead{(arcsec)} & 
    \colhead{} & \colhead{(\unit{cts\perksec})} & \colhead{}\\
  }
\startdata
XMMU~171514$-$-293254 & 258.81098 & -29.54841 & $\pm$1.1$\backslash$0.8 & 4.3 & 2.8$\pm$0.8 &  M1 \\
XMMU~171418$-$-293053 & 258.57689 & -29.51474 & $\pm$0.7$\backslash$0.8 & 7.1 & 3.1$\pm$0.6 &  M2 \\
XMMU~171443$-$-292511 & 258.68223 & -29.41987 & $\pm$0.6$\backslash$0.6 & 5.1 & 2.0$\pm$0.5 &  M3 \\
XMMU~171401$-$-292145 & 258.50750 & -29.36265 & $\pm$1.0$\backslash$0.6 & 6.3 & 3.3$\pm$0.7 &  M4 \\
XMMU~171356$-$-291720 & 258.48705 & -29.28889 & $\pm$0.7$\backslash$1.0 & 4.5 & 3.4$\pm$0.9 &  M5 \\
XMMU~171505$-$-292235 & 258.77451 & -29.37666 & $\pm$1.2$\backslash$0.7 & 5.6 & 3.6$\pm$0.8 &  M6 \\
\enddata
  \label{tab:allMOS}
  \tablecomments{The right ascension (R.A.) and declination (Decl.) are given in decimal degrees.}
  \tablenotetext{a}{Statistical uncertainty from the source detection in seconds of arc.}
\end{deluxetable}


\begin{deluxetable}{cccccccc}
  \tabletypesize{\scriptsize}
  \tablewidth{0pt}
  \tablecaption{\xray\ Catalogue of sources in the field of NGC~6304 
    (Fluxes)}
  \tablehead{
    \colhead{ID} & \colhead{S/N}  & \colhead{Rate\ \tablenotemark{a}} & 
    \colhead{$kT_{\rm eff}$ \tablenotemark{b}} & \colhead{$\alpha$ \tablenotemark{b}} & 
    \colhead{$\chi^2_\nu$/dof (prob.)} & \colhead{$F_{X}$ \tablenotemark{c}} & 
    \colhead{Type \tablenotemark{d}}\\
    \colhead{} & \colhead{}  & \colhead{(\unit{cts\perksec})} & 
    \colhead{(eV)} & \colhead{} & \colhead{} & \colhead{} & \colhead{}\\
  }
  \startdata
 10 &  5.2 &  4.5\ppm1.9 & --             & (1) \tablenotemark{*} & 0.74/16 (0.76) & 0.37 & ??   \\
  9 & 20.2 & 25.2\ppm2.2 & 115\ud{21}{16} & --                    & 1.20/31 (0.21) & 1.52 & qNS? \\
  8 &  5.3 &  7.1\ppm2.0 & --             & -2.0\ud{1.2}{2.2}     & 1.73/7  (0.10) & 2.8  & ??   \\
  7 &  4.3 &  4.3\ppm1.9 & --             & (1) \tablenotemark{*} & 0.84/16 (0.63) & 0.37 & ??   \\
  6 &  8.0 &  7.2\ppm1.7 & --             & (1) \tablenotemark{*} & 0.75/13 (0.71) & 0.76 & ??   \\
  5 &  7.0 &  9.1\ppm2.0 & 70\ud{28}{20}  & --                    & 1.09/16 (0.36) & 0.59 & qNS? \\
  4 & 19.3 & 43.8\ppm2.8 & 122\ud{31}{45} & $1.2\ud{0.7}{0.8}$    & 0.85/42 (0.75) & 2.3  & qNS? \\
  3 &  4.3 &  6.2\ppm1.9 & --             & (1) \tablenotemark{*} & 0.68/16 (0.82) & 0.77 & ??   \\
  2 &  5.4 &  6.1\ppm1.9 & --             & (1) \tablenotemark{*} & 1.37/15 (0.15) & 0.86 & ??   \\
  1 &  8.3 & 15.3\ppm2.1 & --             & 0.52\ppm0.34          & 0.96/15 (0.50) & 2.47 & ??   \\
  0 &  4.7 & 10.3\ppm1.7 & --             & -0.3\ud{0.7}{0.5}     & 0.84/12 (0.60) & 4.2  & ??   \\
  \enddata
  \label{tab:all2}
  \tablecomments{\nh\ column density fixed to $\nhtt=0.266$ for all 
    spectral fits.}
  \tablenotetext{a}{Background subtracted count rate in counts per ks.}
  \tablenotetext{b}{Best fit parameters.  $kT_{\rm eff}$ for a neutron 
    star atmosphere model and $\alpha$ for a power-law component.  
    Both were combined in the case of source \#4.}
  \tablenotetext{c}{Unabsorbed flux (units of $\ee{-13}\cgsflux$) 
    between 0.5--10 \keV.} 
  \tablenotetext{d}{Type of \xray\ source.  qNS = transient neutron star
    in quiescence.}  
  \tablenotetext{*}{The value of the slope in parenthesis are held 
    fixed at the value $\alpha=1$.}
\end{deluxetable}

\begin{deluxetable}{lrlrlcrrc}
  \tablecaption{\label{tab:NS} H-atmosphere spectral parameters for 
    source \#4, \#9 and \#5 for other known probable GC and field
    qNSs}
  \tablewidth{0pt}
  \tabletypesize{\scriptsize}
  \tablehead{
    \colhead{Object} & \colhead{\rinfty} & \colhead{\kteff} & \colhead{d} 
    & \colhead{\nhtt} &  $\alpha$ & \colhead{$L_{X}$} & \colhead{$F_{X}$} & \colhead{Ref.} \\
    \colhead{} & \colhead{(km)} & \colhead{(eV)} & \colhead{(kpc)} & 
    \colhead{} & \colhead{} & \colhead{}& \colhead{} \\
  }
  \startdata
  \sourcenine   & 10.7\ud{6.3}{3.1} & 115\ud{21}{16} & 5.97 & (0.266) & -- & 0.65 & 1.52 \tablenotemark{1}& present work\\
  \sourcefive   & 23\ud{69}{10}     &  70\ud{28}{20} & 5.97 & (0.266) & -- & 0.07 & 0.59 \tablenotemark{1}& present work\\
  \sourcefour   & 11.6\ud{6.3}{4.6} & 122\ud{31}{45} & 5.97 & (0.266) & 1.2\ud{0.7}{0.8}& 1.01 & 2.3 \tablenotemark{1}& present work\\
  \OmCen (ACIS) & 14.3\ud{2.1}{2.1} & 66\ud{4}{5}    & 5    & (0.09)  & -- &  0.5\ppm0.2  & 1.67\ppm0.67 \tablenotemark{2}& 1\\
  \OmCen (EPIC) & 13.6\ud{0.3}{0.3} & 67\ud{2}{2}    & 5.3  & 0.09\ppm0.025 & -- & 0.32\ppm0.02 & 0.95\ppm0.06 \tablenotemark{2}& 2\\
  M13 (EPIC)    & 12.8\ud{0.4}{0.4} & 76\ud{3}{3}    & 7.7  & (0.011) & -- & 0.73\ppm0.06 & 1.03\ppm0.08 \tablenotemark{2}& 3\\
  47 Tuc X5     & 19.0\ud{8.8}{7.8} & 101\ud{21}{14} & 4.85 & 0.09\ud{0.08}{0.05} & -- & 1.4 & 4.3 \tablenotemark{1,*}& 4\\
  47 Tuc X7     & 14.5\ud{1.8}{1.6} & 105.4\ud{5.6}{5.6} & 4.85 & 0.042\ud{0.018}{0.016} & -- & 1.5 & 5.3 \tablenotemark{1,*}& 5\\
  M 28 (\#26)   & 14.5\ud{6.9}{3.8} & 90\ud{30}{10}  & 5.5  & 0.26\ppm0.04 & -- & 1.2\ud{0.7}{0.4} & 3.35\ud{1.9}{1.1} \tablenotemark{3} & 6\\
  M 30 A-1      & 13.4\ud{4.3}{3.6} & 94\ud{17}{12}  & 9.0\ppm0.5 & 0.029\ud{0.017}{0.012} & -- & 0.71 & 0.73  \tablenotemark{1,*}& 7\\
  NGC~6397 (U24)& 4.9\ud{14}{1}     & 57--92          & 2.5   & 0.1--0.26    & -- & 0.08  & 1.06 \tablenotemark{5,*}& 8\\
  M80 CX2       & (10) & 82\ppm{2}      &10.3\ud{0.8}{0.7}& 0.09\ud{0.025}{0}   & -- & 0.29\ppm0.02 & 0.23\ppm{0.02} \tablenotemark{4} & 9\\
  M80 CX6       & (10) & 76\ud{5}{6}    &10.3\ud{0.8}{0.7}& 0.22\ud{0.08}{0.07} & -- & 0.09\ppm0.01 & 0.07\ppm{0.01} \tablenotemark{4} & 9\\
  NGC~2808 C2   & (12) & 81.3\ud{5.3}{4.9} & 9.6 & 0.82\ppm0.40 & -- & 0.26\ppm{0.04} & 0.24\ppm{0.04} \tablenotemark{1} & 10 \\ 
  NGC~3201 16   & (10) & 172\ud{55}{42} & 5.0 & (1.17) & 1.04\ppm{0.63} & 0.34 & 1.0 \tablenotemark{0,*} & 11 \\ 
  \hline
  Aql X-1          & 13.4\ud{5}{4}     & 135\ud{18}{12} & 5    & 0.35\ud{0.08}{0.07} & (1.0) & 4.4 & 14.7 \tablenotemark{1,*} &12\\
  Cen X-4          & 16.8\ud{2.6}{2.6} & 59\ud{6}{6}    & 1.2  & (0.055)             & 1.0\ud{0.6}{0.4}& 0.17 & 0.6 \tablenotemark{1,*,v} &13\\
  $^a$ 4U 2129+47  & 9.3\ud{7.8}{4.6}  & 60\ud{15}{11}  & 1.5  & (0.28)              & -- & 0.07\ud{0.15}{0.03} & 2.4\ud{5.1}{1.0} \tablenotemark{6} &14\\
  $^a$ 4U 2129+47  & 16\ud{11.0}{6.40} & 80\ud{19}{15}  & 6.0  & (0.17)              & -- & 0.56\ud{0.12}{0.3} & 1.3\ud{0.3}{0.7} \tablenotemark{6} &14\\
  $^a$ 4U 1608-522 & 12.3\ud{5.9}{3.5} & 130\ud{20}{20} & 3.6  & (0.8)               & -- & 0.83\ppm0.42 & 5.4\ppm2.7 \tablenotemark{1} &15\\
  $^a$ 4U 1608-522 & 65\ud{47}{26}     & 80\ud{15}{12}  & 3.6  & (1.5)               & -- & 0.73\ppm0.47 & 4.7\ppm3.0 \tablenotemark{1} &15\\
 
  \enddata
  \tablecomments{For the qLMXBs of the present work, the mass was kept 
    fixed to 1.4 \Msun. As mentioned in the text, the tabulated model 
    of the neutron star atmosphere was used for \sourcenine\ and 
    \sourcefive\, while a {\tt nsa+PL} model was used to fit the 
    spectrum of \sourcefour. The top part of the table contains GC 
    qLMXBs while the second part only shows the parameters of field 
    qLMXBs. The inferred radius is measured using the distance 
    to the GC quoted in the text.  Previously measured 
    values of the radius have been converted from $R$ to \rinfty, the 
    projected radius.  Error bars are 90 per cent confidence.  Values in
    parenthesis are held fixed. The unabsorbed \xray\ luminosity $L_{X}$ 
    and flux $F_{X}$ are expressed in units of $\ee{33}\cgslum$ 
    and $\ee{-13}\cgsflux$, respectively, both in \xray\ bands listed 
    below. }
  \tablenotetext{a}{Different assumed values of D and/or \nh.}
  \tablenotetext{0}{0.2--10\keV\ band.}
  \tablenotetext{1}{0.5--10\keV\ band.}
  \tablenotetext{2}{0.1--5\keV\ band.}
  \tablenotetext{3}{0.5--8\keV\ band.}
  \tablenotetext{4}{0.5--6\keV\ band.}
  \tablenotetext{5}{0.5--2.5\keV\ band.}
  \tablenotetext{6}{0.5--3.0\keV\ band.}
  \tablenotetext{*}{No error was provided by the authors.}
  \tablenotetext{v}{Indicates that the source flux varies.}
  \tablecomments{References: 
    1, \cite{rutledge02b};  
    2, \cite{gendre03a};
    3, \cite{gendre03b};
    4, \cite{heinke03};
    5, \cite{heinke06};
    6, \cite{becker03};
    7, \cite{lugger07};
    8, \cite{grindlay01b};
    9, \cite{heinke03c};
    10, \cite{servillat08};
    11, \cite{webb06}
    12, \cite{rutledge01b};  
    13, \cite{rutledge01};
    14, \cite{rutledge00};
    15, \cite{rutledge99}
  }
\end{deluxetable}


\begin{deluxetable}{ccc|ccc}
  \tabletypesize{\scriptsize}
  \tablecaption{\label{tab:CompFluxB} Difference in the Luminosities 
    between the \rosat\ and the current \xmm\ observations}
  \tablewidth{0pt}
  \tablehead{
    \multicolumn{4}{c}{XMM-pn} &\multicolumn{2}{c}{ROSAT} \\
    \colhead{ID} & \colhead{$kT_{\rm eff,BB}$}\tablenotemark{a} 
    & ECF \tablenotemark{b} & \colhead{${\rm ID_{\rm R94}}$} 
    & \colhead{Counts predicted\tablenotemark{c}}  
    & \colhead{Counts Observed\tablenotemark{d}} }
  \startdata
  4 & 260\ppm30 & 0.0935\ppm0.0015 &  A &20.7\ppm1.5  & 15  \\
  5 & 170\ppm30 & 0.0855\ppm0.0065 &   B &3.9\ppm0.9   & 7   \\
  9 & 200\ppm30 & 0.090\ppm0.002  & C & 11.4\ppm1.0     & 17  \\
  - & --    & 0.062   &  D  &$<$1.8  & 14  \\
  \enddata
  \tablecomments{For source D, due to the non detection, a $3\sigma$ 
    upper limit is calculated.}
  \tablenotetext{a}{ The $kT_{\rm eff, BB}$ is the black-body effective
    temperature, from spectral fits to the present XMM data.}
  \tablenotetext{b}{ The energy-conversion-factor (ECF) is the number of
    ROSAT/HRI counts per XMM count for each source, which depends on
    $kT_{\rm eff, BB}$ and \nh\ (assumed to be $\nhtt=0.266$). For R94 
    source D, undetected in the present observation, an absorbed thermal 
    bremsstrahlung spectrum with $kT=3\keV$ is assumed.}
  \tablenotetext{c}{The total counts predicted for the R94 observation
    based on the present XMM observation, using the source spectrum
    measured here; the predicted HRI countrate from WebPIMMS; and the
    ROSAT/HRI observation duration of 5030 sec.}
  \tablenotetext{d}{The total number of source counts from ROSAT/HRI
    observation (R94), corrected for observation at the HRI centre
    (that is, after subtracting background counts, correcting for
    vignetting, quantum efficiency and scattering).  The uncertainty 
    in this number is not given by R94.}
\end{deluxetable}


\begin{deluxetable}{lrlrl}
  \tablecaption{\label{tab:Rcore} Distance from the core of the 
    candidate qLMXBs and other known or probable qNSs in GCs.}
  \tabletypesize{\scriptsize}
  \tablewidth{0pt}
  \tablehead{
    \colhead{Object Name} & \colhead{$d_{\rm c}$} & 
    \colhead{$r_{\rm c}$} & \colhead{$\Delta/r_{\rm c}$} & 
    \colhead{Ref.} \\
    \colhead{} & \colhead{(arcmin)}  & \colhead{(arcmin)} & 
    \colhead{} & \colhead{}\\
  } 
  \startdata
  \sourcenine            &  6.7  & 0.21 &31.8  & 0\\
  \sourcefour            &  0.17 & 0.21 & 0.79 & 0\\
  \sourcefive            &  3.2  & 0.21 &15.2  & 0\\
  $\omega$Cen            &  4.38 & 2.58 & 1.7  & 1\\
  M13                    &  0.74 & 0.78 & 0.95 & 2,10\\ 
  47 Tuc X7              &$<0.37$& 0.38 &$<1.5$& 3 \\
  47 Tuc X5              &$<0.37$& 0.38 &$<1.5$& 3 \\ 
  M 28 (\#26)            &  0.05 & 0.24 & 0.21 & 4,10\\
  M 30 A-1               &  0.03 & 0.06 & 0.5  & 5,10\\
  NGC~6397 (U24) \tablenotemark{cc} &  0.34 & 0.05 & 6.8  & 6,10\\
  M80 CX2                & 0.063 & 0.15 & 0.42 & 7,10\\
  M80 CX6                & 0.324 & 0.15 & 2.2  & 7,10\\
  NGC~2808 C2            & 0.08  & 0.26 & 0.31 & 8,10\\
  NGC~3201 16            & 5.8   & 1.43 & 4.1  & 9,10\\
  \enddata
  \tablecomments{This table is a compilation of the positional information 
    of known qLMXBs and candidates qLMXBs in GCs.  The columns, from left 
    to right, are the object name (or the GC hosting the qLMXB), the 
    distance of the qLMXB from the optical centre of the GC (in arcminutes),
    the core radius of the GC (in arcminutes) and the distance of the qLMXB
    from the optical centre of the GC in units of core radii.\\
    References : 0, present work;
    1, \cite{rutledge02b};
    2, \cite{gendre03b};
    3, \cite{heinke03};
    4, \cite{becker03};
    5, \cite{lugger07};
    6, \cite{grindlay01b};
    7, \cite{heinke03c};
    8, \cite{servillat08};
    9, \cite{webb06};
    10, \cite{harris96}, for the core radii values.}
  \tablenotetext{cc}{NGC~6397 is a core collapsed star cluster.}
\end{deluxetable}

\clearpage

\end{document}